\documentclass[conference]{IEEEtran}
\IEEEoverridecommandlockouts

\usepackage{cite}
\usepackage{amsmath,amssymb,amsfonts}
\usepackage{algorithmic}
\usepackage{textcomp}
\usepackage{fancyhdr}
\usepackage[hyphens]{url}
\usepackage[normalem]{ulem}
\usepackage[utf8]{inputenc}
\usepackage{enumitem}

\usepackage{xcolor}
\usepackage{amssymb}
\usepackage{graphicx}
\usepackage{subfig}
\usepackage{listings}
\usepackage[linesnumbered,ruled,vlined]{algorithm2e}
\usepackage[frozencache=true,cachedir=.]{minted}
\usepackage{pifont}
\usepackage{multirow}
\usepackage{soulutf8}
\soulregister\Hl{7}
\soulregister\ref7
\soulregister\cite7
\soulregister\pageref7

\renewcommand\hl[1]{#1}

\SetCommentSty{mycommfont}

\SetKwInput{KwInput}{Input}                
\SetKwInput{KwOutput}{Output}              

\def\BibTeX{{\rm B\kern-.05em{\sc i\kern-.025em b}\kern-.08em
    T\kern-.1667em\lower.7ex\hbox{E}\kern-.125emX}}
\begin{document}

\title{Dual-side Sparse Tensor Core \\
\thanks{\IEEEauthorrefmark{1} Contribution during internship at Microsoft Research}
\thanks{\IEEEauthorrefmark{7} Corresponding Author}}

\author{\IEEEauthorblockN{
Yang Wang\IEEEauthorrefmark{1}\IEEEauthorrefmark{2}\IEEEauthorrefmark{3}, 
Chen Zhang\IEEEauthorrefmark{7}\IEEEauthorrefmark{3}, 
Zhiqiang Xie\IEEEauthorrefmark{1}\IEEEauthorrefmark{4}, 
Cong Guo\IEEEauthorrefmark{1}\IEEEauthorrefmark{5}, 
Yunxin Liu\IEEEauthorrefmark{3}, 
Jingwen Leng\IEEEauthorrefmark{5}}

\IEEEauthorblockA{
\textit{
\IEEEauthorrefmark{2} University of Electronic Science and Technology of China, \IEEEauthorrefmark{3} Microsoft Research} \\
\textit{
\IEEEauthorrefmark{4} ShanghaiTech University,
\IEEEauthorrefmark{5} Shanghai Jiao Tong University} \\
\textit{\{t-yangwa, zhac, yunxin.liu\}@microsoft.com} \\ \textit{xiezhq@shanghaitech.edu.cn, \{guocong, leng-jw\}@sjtu.edu.cn}
}
}

\maketitle


\begin{abstract}

Leveraging sparsity in deep neural network (DNN) models is promising for accelerating model inference. Yet existing GPUs can only leverage the sparsity from weights but not activations, which are dynamic, unpredictable, and hence challenging to exploit. 
In this work, we propose a novel architecture to efficiently harness the dual-side sparsity (i.e., weight and activation sparsity). We take a systematic approach to understand the (dis)advantages of previous sparsity-related architectures and propose a novel, unexplored paradigm that combines outer-product computation primitive and bitmap-based encoding format.
We demonstrate the feasibility of our design with minimal changes to the existing production-scale inner-product-based Tensor Core.
We propose a set of novel ISA extensions and co-design the matrix-matrix multiplication and convolution algorithms, which are the two dominant computation patterns in today's DNN models, to exploit our new dual-side sparse Tensor Core.
Our evaluation shows that our design can fully unleash the dual-side DNN sparsity and improve the performance by up to one order of magnitude with \hl{small} hardware overhead.
\end{abstract}



 


\section{Introduction}

As deep learning is widely deployed, there is an emerging need to support billions of queries of inference per day, which is rapidly outpacing training in data centers\cite{hazelwood2018applied}.
Especially, many AI applications have a stringent constraint of service level agreement. Running models at high-scale, low-latency, and high energy efficiency has always been extremely desirable. Model compression and sparsification have become critical optimizations to reduce the number of parameters as well as arithmetic operations and to improve the computational and energy efficiency on various hardware platforms, including ASICs~\cite{sparch,sparten,scnn,eyeriss, cnvlutin, cambriconx, cambricons, hegde2019extensor, eie}, GPUs~\cite{aaai19,dong2017more,figurnov2017spatially,yao2014gpu,sma_dac20}, and FPGAs~\cite{fpga19,han2017ese, lu2018spwa, aimar2018nullhop,caffeine}. 

Realizing the acceleration potential of sparse neural networks, GPU vendors have introduced architectural support to exploit this opportunity.
In particular, sparse Tensor Core~\cite{sparsetensor, a100, mishra2021accelerating} is newly invented to leverage the weight sparsity in DNN models. 
The latest NVIDIA Ampere architecture~\cite{a100}\cite{mishra2021accelerating} introduces a new sparse Tensor Core design 
with a fixed 50\% weight pruning target and achieves a better accuracy and performance trade-off~\cite{fpga19, aaai19}.



Besides the weight sparsity, DNN models also exhibit another form of sparsity called \emph{activation} sparsity, which is introduced by activation functions~\cite{agarap2018deep} and is widely embedded in activation feature maps, for both computer vision~\cite{shi2017speeding} and natural language processing~\cite{DBLP:journals/corr/abs-1810-04805} tasks.
Many previous works have reported a high activation sparsity ranging from 50\% to 98\%~\cite{scnn, cnvlutin}.
However, the current sparse Tensor Core is only able to take advantage of weight sparsity but not activation sparsity.
How to effectively leverage the activation sparsity remains an open and challenging research problem, because the activation sparsity dynamically changes with the input and cannot be pre-determined and controlled by the pruning method such as the vector-based pruning\cite{sparsetensor}.

Although there are prior efforts that tackle the dual-side sparsity in the context of ASIC designs, they cannot be directly adopted by GPUs.
The wide applicability of GPUs requires the support of both sparse general matrix-matrix multiplication (SpGEMM) and sparse convolution (SpCONV).
Those two are key computation kernels in today's DNN models ranging from convolutional neural networks (CNNs)~\cite{resnet, DBLP:journals/corr/SimonyanZ14a}, recurrent neural networks (RNN)~\cite{lstm, DBLP:conf/emnlp/LuongPM15} to attention-based neural networks~\cite{DBLP:journals/corr/abs-1810-04805,guan2020far}.
Unfortunately, current ASIC designs only consider SpGEMM kernel~\cite{outerspace, sparch, MatRaptor} or SpCONV kernel~\cite{scnn,extensor,sparten}.
In this work, we aim to accelerate \textit{both} dual-side SpCONV and SpGEMM on Tensor Core.

\begin{table}[t!]
\centering
\caption{\hl{Technical differences to related work.}}
\label{tab:tech_diff}
\begin{tabular}{|c|c|c|c|}
\hline
                                                                   & \begin{tabular}[c]{@{}c@{}}Inner-product\end{tabular} & \begin{tabular}[c]{@{}c@{}}Outer-product\end{tabular} & Misc \\ \hline
\begin{tabular}[c]{@{}c@{}}CSR\end{tabular}     &     \cite{sparsetensor,extensor}     &    \cite{outerspace,sparch,scnn}   & \cite{MatRaptor}    \\ \hline
\begin{tabular}[c]{@{}c@{}}Bitmap\end{tabular} &     \cite{sparten}   &         \textbf{Our work}          & -  \\ \hline
\end{tabular}
\vspace*{-0.4cm}
\end{table}

The greatest challenge of supporting the dual-side SpCONV and SpGEMM on Tensor Core is the unpredictable and randomly distributed non-zero elements inside the input tensors.
The dot product unit, a basic computational unit in Tensor Core hardware, conducts a vector-vector inner product. Sparse Tensor Core~\cite{a100}\cite{mishra2021accelerating} resolves the irregularity of weight sparsity by applying a structural pruning scheme, which enforces a constant 50\% weight sparsity to balance the workload and to exploit parallelism in the dot-product unit. However, this method cannot be applied to SpGEMM because activation sparsity is input-dependent and cannot be pre-determined by pruning methods. \hl{Prior ASIC designs take two approaches to leverage dual-side sparsity, as is summarized in Table~\ref{tab:tech_diff}.} SparTen and Extensor~\cite{sparten,extensor} accelerate inner-product by designing dedicated hardware for the inner joint process, which figures out non-zero elements by matching positions in two sparse vectors and accessing those elements. But their method introduces considerable overhead, including complex prefix sum hardware and explicit barrier as the number of non-zeros to be matched is unpredictable. \hl{The other type of work (e.g., OuterSPACE\cite{outerspace} and SpArch\cite{sparch}) accelerate SpGEMM with outer-product, but they do not design for SpCONV, which incurs great performance overhead to transform SpCONV to SpGEMM straightforwardly. Moreover, they are targeted at matrix density $6 \times 10^{-3}$ to $5 \times 10^{-5}$, which become inefficient for mainstream DNN models with the typical density range of $5 \times 10^{-1}$ to $1 \times 10^{-2}$. Likewise, SCNN\cite{scnn} only considers SpCONV but does not support SpGEMM.} 


It's also challenging to accelerate SpCONV. 
GPUs usually transform a CONV operator into a GEMM operator via the im2col method. 
\hl{Sparse Tensor Core~\cite{sparsetensor,a100,mishra2021accelerating} only leverages weight sparsity and the input remains dense, which requires only dense im2col.
However, to leverage dual-side sparsity, the convolution computation now must consider the \emph{sparse} im2col, which slides over the sparse input tensor and incurs irregular memory accesses.}
Moreover, because of the space and time overhead of performing an explicit im2col, vendor-supplied DNN acceleration library~cuDNN\cite{chetlur2014cudnn} provides optimization of \textit{implicit im2col}.
The implicit method fuses the address generation process of im2col into matrix multiplication, and has the best performance in general cases.\footnote{Although there are other faster convolution methods~\cite{lavin2016fast} than implicit im2col, they only prevail on specific matrix shapes, like Winograd for $3\times 3$ convolution kernel. In this work, we only focus on implicit im2col, which outperforms other methods on the majority cases.}
However, performing the implicit im2col on the sparse input tensors is significantly more challenging than on the dense tensors because of the randomly distributed non-zero elements.
In fact, we show that a naive implementation of implicit sparse im2col can be 10$\times$ to 100$\times$ slower than its dense version.

To tackle the problems above, we analyze the computation patterns of sparse im2col and SpGEMM and compare combinations of different approaches. We argue that bitmap-based encoding format is more friendly for efficient sparse im2col acceleration and outer-product is more efficient to leverage SpGEMM's opportunities on Tensor Core, as is summarized in Table~\ref{tab:tech_diff}. We thereby propose a bitmap-based sparse im2col algorithm for SpCONV and an outer-product-based dual-side sparse Tensor Core architecture for SpGEMM. To further co-optimize sparse im2col and SpGEMM, we propose an outer-product-friendly sparse im2col method and a bitmap-based outer-product SpGEMM algorithm. Combining the above techniques, we achieve an efficient implicit sparse im2col design for SpCONV acceleration. Verified on Accel-Sim with V100 architecture, we demonstrate efficient acceleration of both SpCONV and SpGEMM on the proposed dual-side sparse Tensor Core architecture, achieving a speedup of up to one order of magnitude compared with state-of-the-art baselines and imposing negligible hardware overhead. 

The key technical contributions of this work are as follows:

\begin{itemize}[leftmargin=*]
    \item We propose a novel method that combines outer product and bitmap encoding to accelerate SpGEMM (Section~\ref{sec:spgemm}) and SpCONV (Section~\ref{sec:conv}). \hl{To the best of our knowledge, our work is the first to study the sparse and implicit im2col method that is critical for SpCONV acceleration on GPUs.} 
    \item \hl{We show the architectural friendliness of our method to existing GPUs by proposing a small set of modifications, which transform the existing Tensor Core to harness the dual-side sparsity. We also propose novel instruction set extensions that let us leverage the existing high-performance libraries to accelerate SpGEMM and SpCONV (Section~\ref{sec:arch}).}
    \item Through extensive evaluations,
    our dual-side sparse Tensor Core achieves a speedup of up to 7.49$\times$ for SpCONV and up to 8.45$\times$ for SpGEMM over state-of-the-art methods, with a \hl{small} 1.5\% area overhead. (Section~\ref{sec:exp})
\end{itemize}



\section{Background and Related Work}


\subsection{Opportunities of sparsity in DNNs}

\emph{Weight sparsity} has been extensively explored in many prior arts, including computer vision and natural language processing tasks~\cite{han2015learning,liu2015sparse,wen2016learning,ioannou2017deep,yang2019energyconstrained,wavernn,tw_sc20,9157343,yang2019ecc}. They demonstrate high sparsity with various pruning methods. However, a significant reduction in weights can only save storage costs, but hardly
speed up inference due to the fragmented irregular pattern of the pruning method~\cite{tw_sc20}. Some researchers~\cite{yu2017scalpel,wen2016learning,liu2018efficient,sparsetensor,varma2019dynamic,tw_sc20} achieve practical speedup by proposing hardware friendly pruning methods. In NVIDIA's latest Ampere GPU, sparse Tensor Core is first introduced in GPU architecture by adopting the fine-grained structural pruning\cite{aaai19,fpga19,a100,mishra2021accelerating}.

\emph{Activation sparsity} naturally occurs in CNN's feature maps and RNN's hidden layers, followed by ReLU~\cite{agarap2018deep} activation functions. Different from weight sparsity that can apply structural pruning by artificial efforts, activation sparsity dynamically changes with input images and is featured with a highly unstructured pattern. Previous works~\cite{shi2017speeding,scnn,seernet} demonstrate that activation sparsity can be as high as 45\% to 98\%. Some researchers\cite{ren2018sbnet,kong2017ron} accelerate activation sparsity on GPUs by adding blocked masks, but it requires external knowledge and is not generic. 
Besides performance acceleration, prior works have also exploited the DNN sparsity to improve their robustness~\cite{path_cvpr19,ptolemy_micro20}.
However, to the best of our knowledge, no previous work has demonstrated meaningful speedup by exploiting activation sparsity on GPU.

\subsection{Computation kernels}

Deep neural networks are composed of multiple structurally connected layers of linear and non-linear functions. Among them, matrix multiplication and convolution are the major computation kernels with dominating amount of parameters and computation workloads\cite{cong2014minimizing,caffeine}.

Matrix multiplication (GEMM) is the major computation kernel in NLP models, e.g., RNNs\cite{wavernn} and attention-based models\cite{DBLP:journals/corr/abs-1810-04805}. 
Dense GEMM is one of the fundamental computation primitives provided by GPU, which has been under continuous optimization. Especially, Tensor Core, as the specialized hardware, has recently been deployed in GPU to boost GEMM performance by an order of magnitude. 

Convolution has played a key role in CNNs\cite{resnet,ren2018sbnet}, which often takes over more than 90\% of the overall workload\cite{cong2014minimizing}. The convolution takes in a number of $C$ input feature maps, each sized of $H\times W$. Every input feature map is convolved by a sliding kernel sized of $K\times K$ to calculate one pixel in the output feature map. A total of $N$ feature maps will be generated as output to the next layer.

\begin{figure}[t]
 \centering
  \includegraphics[width=1\linewidth]{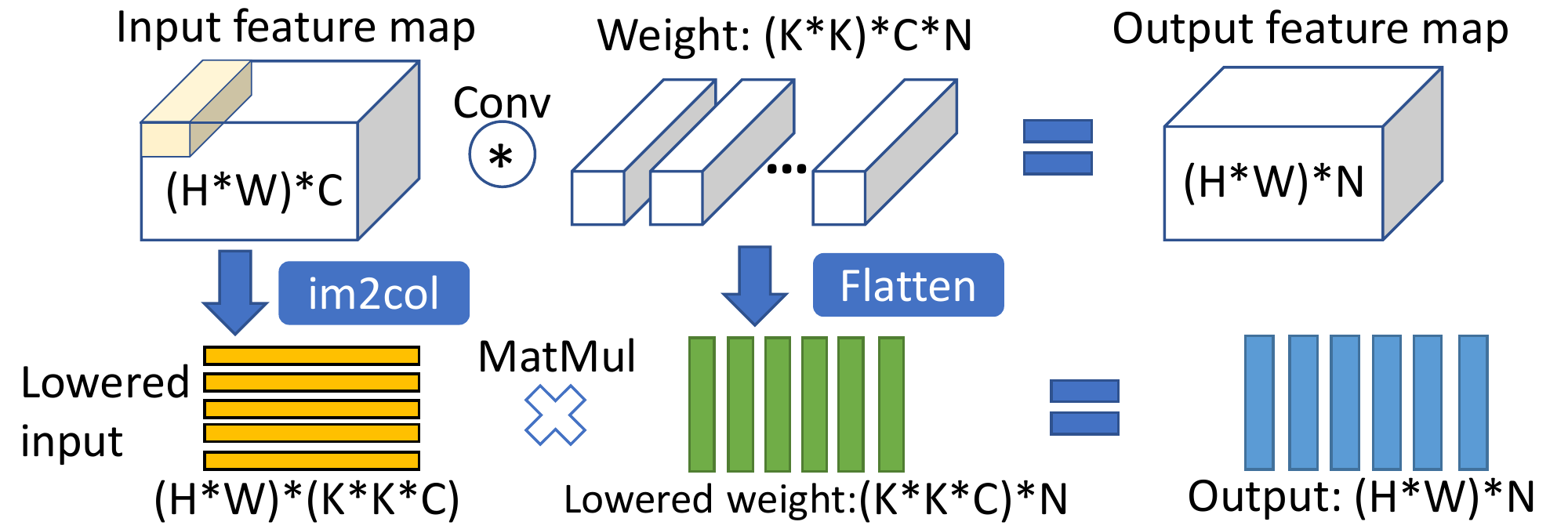}
  \caption{\small The im2col-based transformation of CONV to GEMM.} 
  \vspace{-0.1in}
  \label{fig:im2col_conv}
\end{figure}

To leverage tensor core's matrix-multiply primitives, state-of-the-art DNN acceleration libraries, e.g., cuDNN, usually transforms convolution into matrix-multiplication by applying im2col function.  Figure~\ref{fig:im2col_conv} shows an overview of the im2col function. It re-arranges and expands convolution's input feature maps into a matrix, called lowered feature map, whose each row corresponds to a location of 2-dim sliding window in convolution's input feature maps. Im2col on weight parameters is simply flattening each $K\times K\times C$ kernel. Since im2col's data expansion is mainly applied on input feature maps, the weight-sparse architecture treats im2col as \textit{dense im2col} while the dual-side sparse architecture has to deal with \textit{sparse im2col}, which has not been fully discussed in previous work. 

The na\"ive approach, called \textit{explicit im2col}, conducts im2col and GEMM separately. One concern about \textit{explicit im2col} is that the lowered feature map usually takes $K\times K$ times more global memory than the original feature map because the overlapped sliding windows generate duplicated data. To improve the input data reuse, the state-of-the-art DNN library (e.g., cuDNN) uses \textit{implicit im2col}, which keeps original feature map layout in global memory and uses an address conversion scheme to do im2col transformation in on-chip caches, instead of physically duplicating data in global memory. Implicit im2col has been widely used as the state-of-the-art method in accelerating convolution with GEMM operators. 

\subsection{Design philosophy and challenges}

\hl{
In this paper, our design target is to transform the existing Tensor Core with minimal modification to exploit both weight and activation sparsity for better performance.
The alternative design is to directly port an existing ASIC as the GPU's co-processor. 
However, this design philosophy leads to a significant area cost. 
E.g., SpArch\cite{sparch} uses an array with 16 floating point multipliers to compute the outer-product for SpGEMM, and requires specialized Merge Tree and matrix read/write hardware, which occupies the 98.4\% die area ($\approx 28mm^2$ @ 40nm). 
If we were to scale the design to 40960 FP multipliers in V100~\cite{v100}, the estimated area cost would be prohibitively expensive (i.e., $71680~mm^2$).
Instead, our design tries to reuse the GPU's hardware resources such as memory hierarchies and data path, which leads to a small overhead ($\approx 12.8 mm^2$, or 1.5\% as we show later).
}

\hl{
The challenges for our design philosophy lie in the aspect that we need to support both the SpGEMM and SpCONV.
The prior SpGEMM designs such as OuterSPACE\cite{outerspace} and SpArch\cite{sparch} target at matrix density $6 \times 10^{-3}$ to $5 \times 10^{-5}$, which become inefficient for mainstream DNN models with the typical density range of $5 \times 10^{-1}$ to $1 \times 10^{-2}$. 
Moreover, those designs do not support SpCONV, where the aforementioned im2col approach imposes a unique challenge as it needs to handle the sparse and unpredictable non-zero elements in the input tensor.
As such, none of the prior designs serve our purpose, and we need new architectural innovations.
}

\section{Bitmap-based SpGEMM}
\label{sec:spgemm}

\begin{figure}[t!]
\centering
\subfloat[][Dense outer product.]{\includegraphics[width=0.85\linewidth]{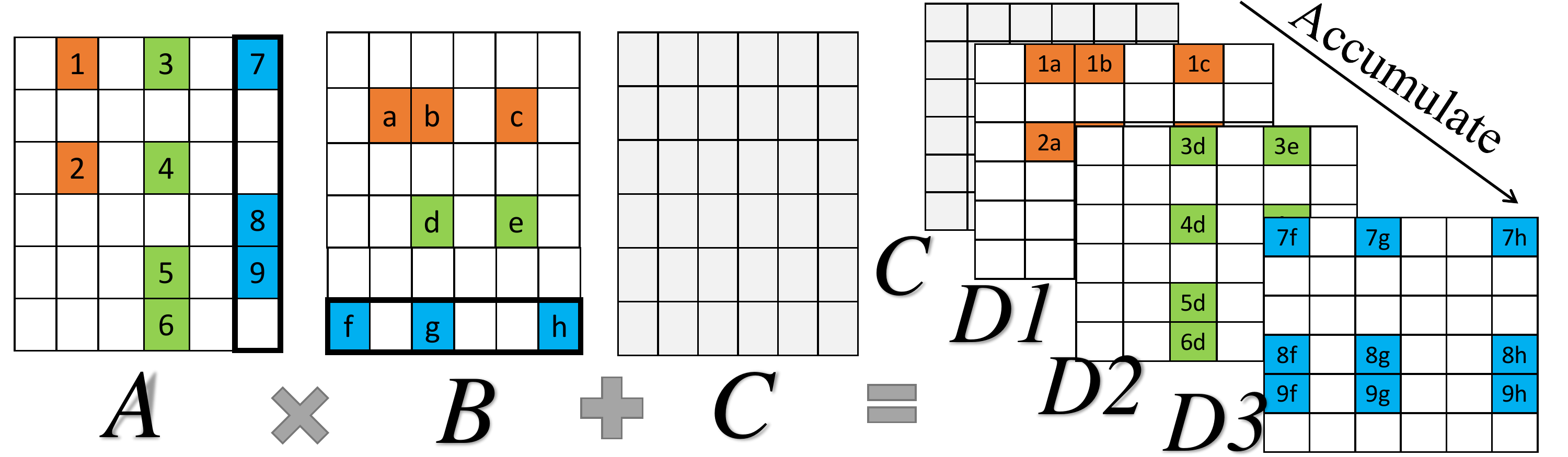}\label{fig:algo_overview_1}}\\ 
\subfloat[][Encoded bitmaps of Matrix A and B.]{\includegraphics[width=0.73\linewidth]{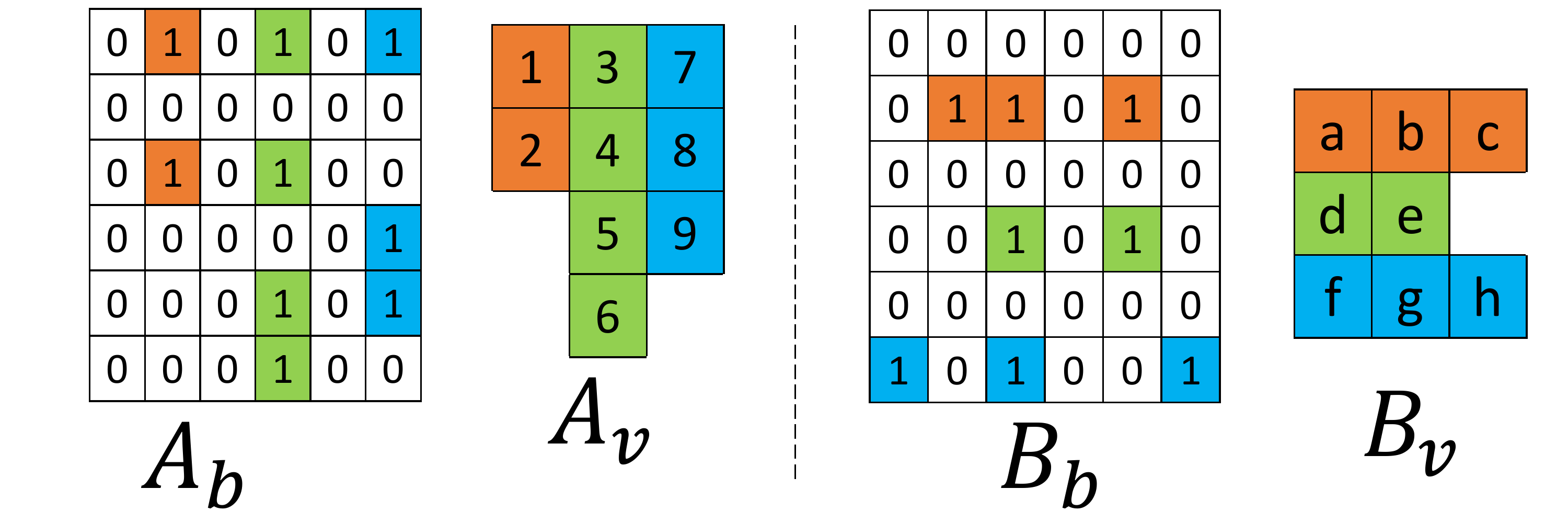}\label{fig:algo_overview_2}}\\
\subfloat[][\hl{Procedures of the proposed SpGEMM.}]{
\includegraphics[width=0.97\linewidth]{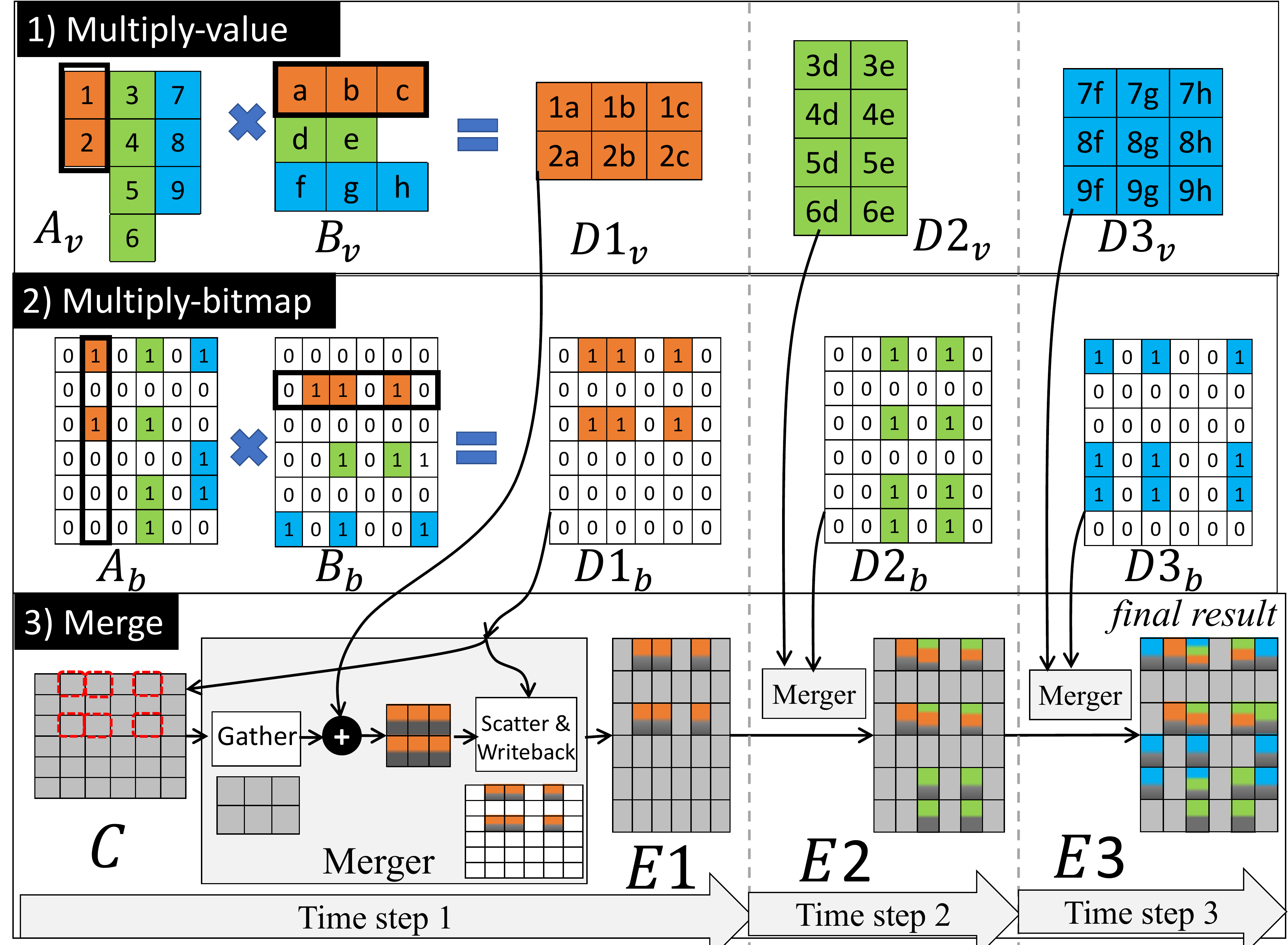}\label{fig:algo_overview_3}
}
\caption{\small Our proposed bitmap-based outer-product SpGEMM.} 
\label{fig:algo_overview}
\end{figure}

We propose an outer-product-based algorithm to accelerate SpGEMM using the bitmap-based sparse encoding format. 



\subsection{Overview}
To exploit the dual-side sparsity, we propose an efficient SpGEMM algorithm based on outer-product matrix multiplication. A basic step in outer-product-based matrix multiplication is to compute a cross product between a column of $\mathbb{A}$ (sized of $M\times 1$) and a row of $\mathbb{B}$ (sized of $1\times N$), which leads to the output $M\times N$ partial matrix (e.g., $\mathbb{D}1,~\mathbb{D}2,~\mathbb{D}3$ in Figure~\ref{fig:algo_overview_1}). 
To generate the final output, we need to accumulate all those partial results and bias matrix $\mathbb{C}$ with multiple rounds. 


Our approach achieves an efficient outer product by using bitmap representation, as shown in Figure~\ref{fig:algo_overview_2}. Each input matrix is represented by a two-tuple encoding of a bitmap (e.g., $\mathbb{A}_b$ and $\mathbb{B}_b$), and a collection of non-zero values (e.g., $\mathbb{A}_v$ and $\mathbb{B}_v$). The bitmap uses 1's for positions of non-zero values and 0's for zeros. To support outer-product, matrix $\mathbb{A}_v$ is encoded in column-major and $\mathbb{B}_v$ is encoded in row-major.

The proposed SpGEMM algorithm has three major operations on the bitmap encoded matrices, which are \textit{multiply-value}, \textit{multiply-bitmap} and \textit{merge} respectively. In Figure~\ref{fig:algo_overview_3}, the \textit{multiply-value} operation computes the cross-product on each vector-vector pair of $\mathbb{A}_v$ and $\mathbb{B}_v$ to generate values of the partial matrices (e.g., $\mathbb{D}1_v$, $\mathbb{D}2_v$, and $\mathbb{D}3_v$). Since the outer-product avoids the explicit inner-join process, its multiplication is regular and easy to accelerate. The \textit{multiply-bitmap} operation computes 1-bit cross product on the bitmaps $\mathbb{A}_b$ and $\mathbb{B}_b$. The output bitmap contains the sparsity information of the corresponding partial matrix, such as $\mathbb{D}1_b$ to $\mathbb{D}3_b$. At last, the \textit{merge} operation uses values (e.g., $\mathbb{D}1_v$) and bitmaps (e.g., $\mathbb{D}1_b$) of the partial matrices from the previous two operations to do the accumulation with multiple rounds, such as from $\mathbb{E}1$ to $\mathbb{E}3$. Despite the benefits from the regular multiplication provided by the outer product, the partial matrix is sparse and irregular. However, it is worthwhile to make this trade-off because dealing with a single-side irregular accumulation is much cheaper than dual-side sparse multiplication. We propose a gather-scatter method to merge the non-zero values from different partial matrices with multiple rounds.

In the following section, we present a detailed analysis on the problems of inner-product-based Tensor Core with accelerating SpGEMM and the advantages of outer-product-based Tensor Core. Then we propose a SpGEMM algorithm for the outer-product Tensor Cores in a warp. At last, we extend to the whole device.

\begin{figure}[t]
\centering
\subfloat[][]{\includegraphics[width=0.29\linewidth]{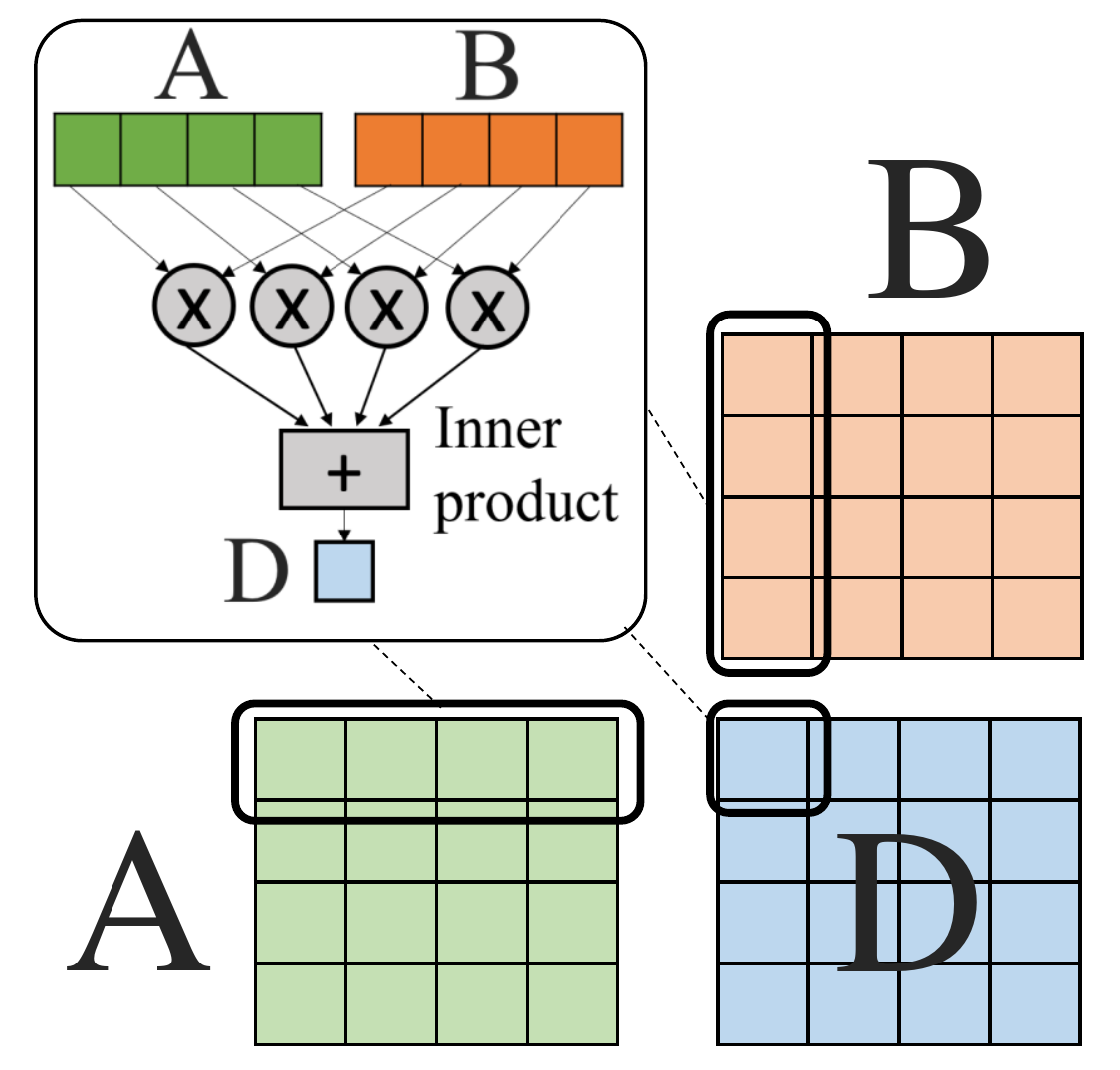}\label{fig:inner_prod_prob_1}}
\subfloat[][]{\includegraphics[width=0.34\linewidth]{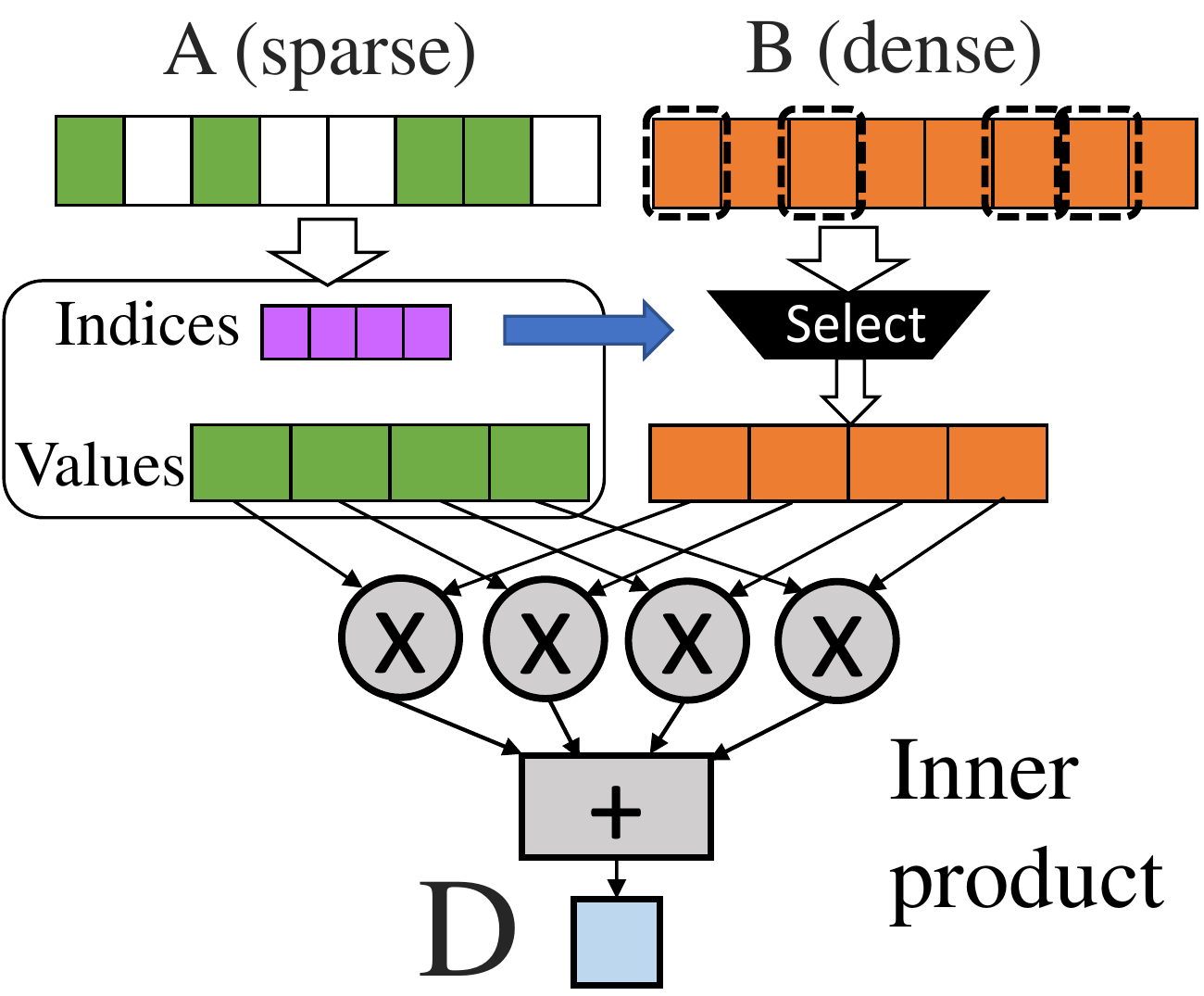}\label{fig:inner_prod_prob_2}}
\subfloat[][]{
\includegraphics[width=0.34\linewidth]{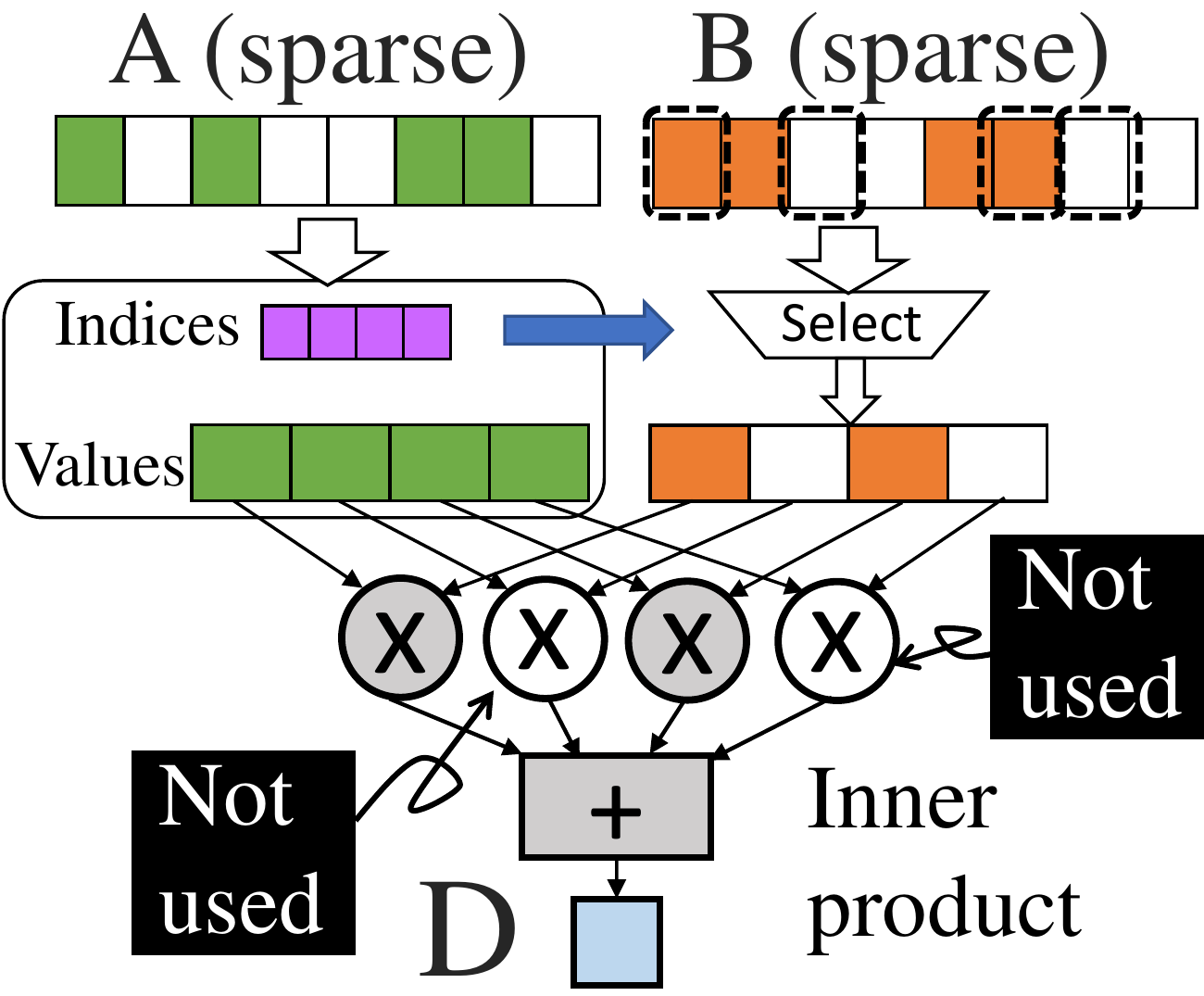}\label{fig:inner_prod_prob_3}
}
\caption{\small (a) 4$\times$4$\times$4 matrix multiplication primitive used in inner-product-based Tensor Core \cite{v100}; (b) \hl{A sparse inner-product unit in A100\cite{a100}\cite{mishra2021accelerating}}; \hl{(c) Dual-side sparsity unit based on inner-product.}}
\label{fig:wmma_outer_prod}
\end{figure}


\begin{figure}[t]
\centering
\subfloat[][]{\includegraphics[width=0.25\linewidth]{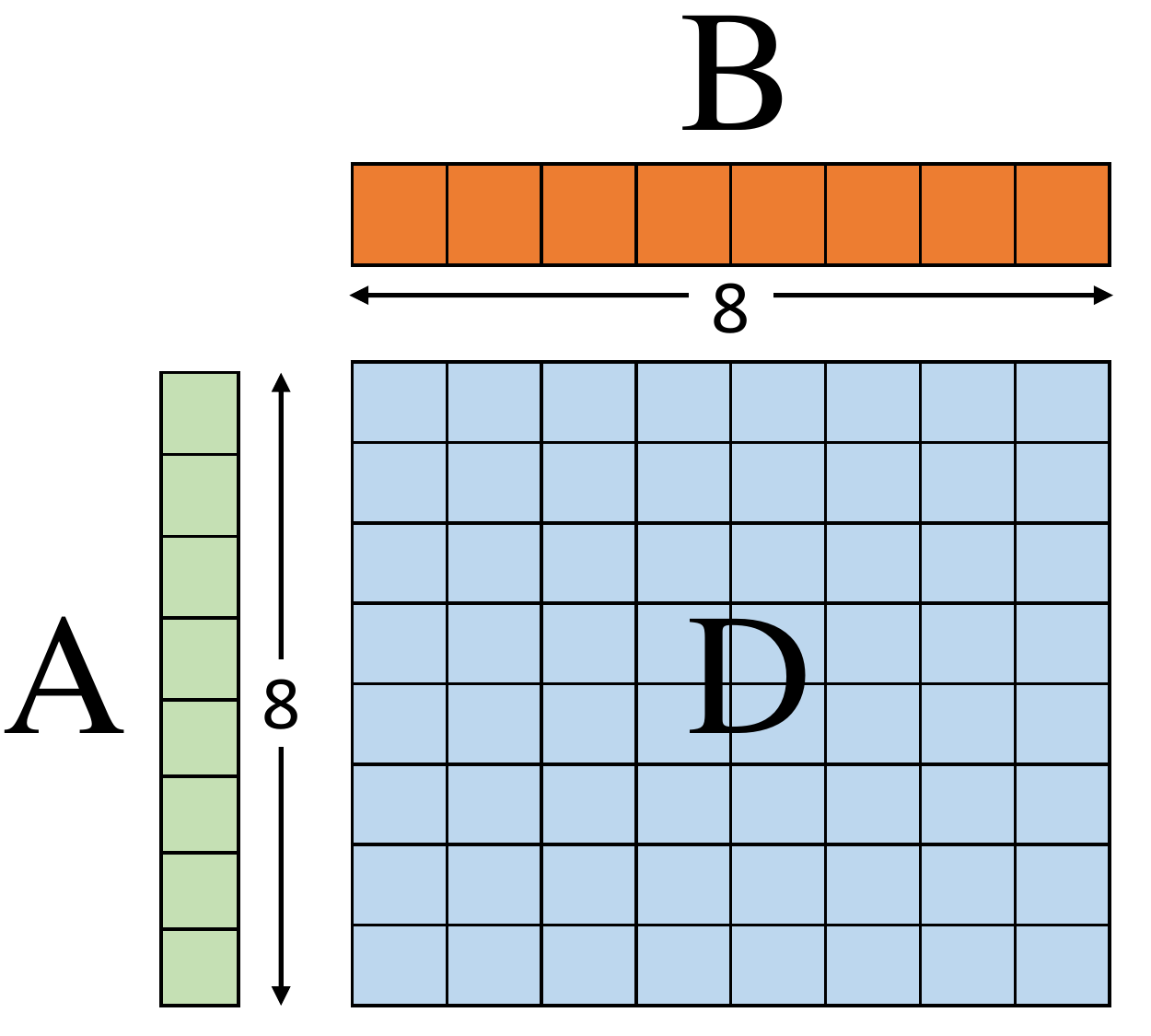}
\label{fig:outer_prod_prob_0}}
\subfloat[][]{
\includegraphics[width=0.25\linewidth]{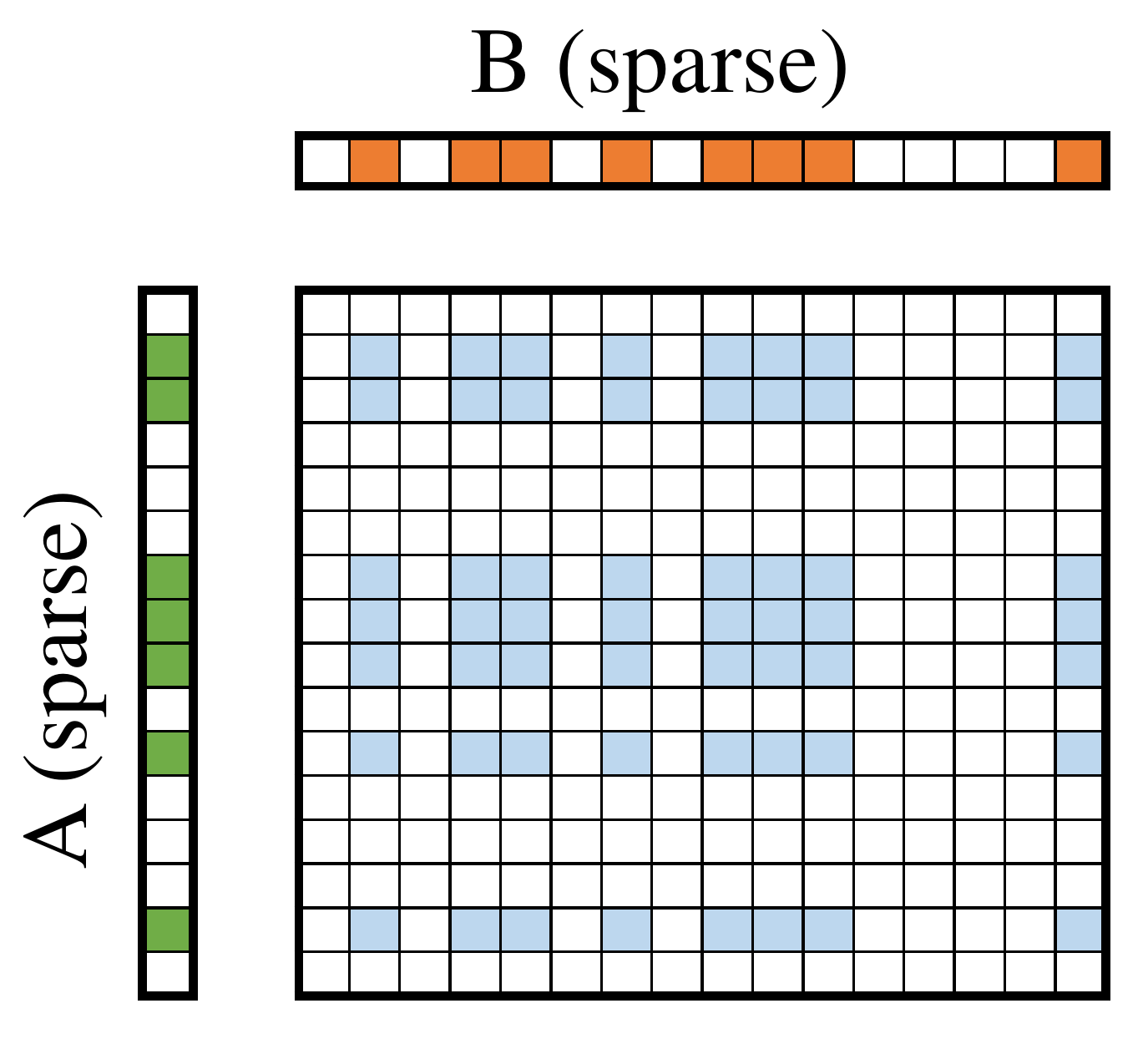}\label{fig:outer_prod_prob_1}
}
\subfloat[][]{\includegraphics[width=0.35\linewidth]{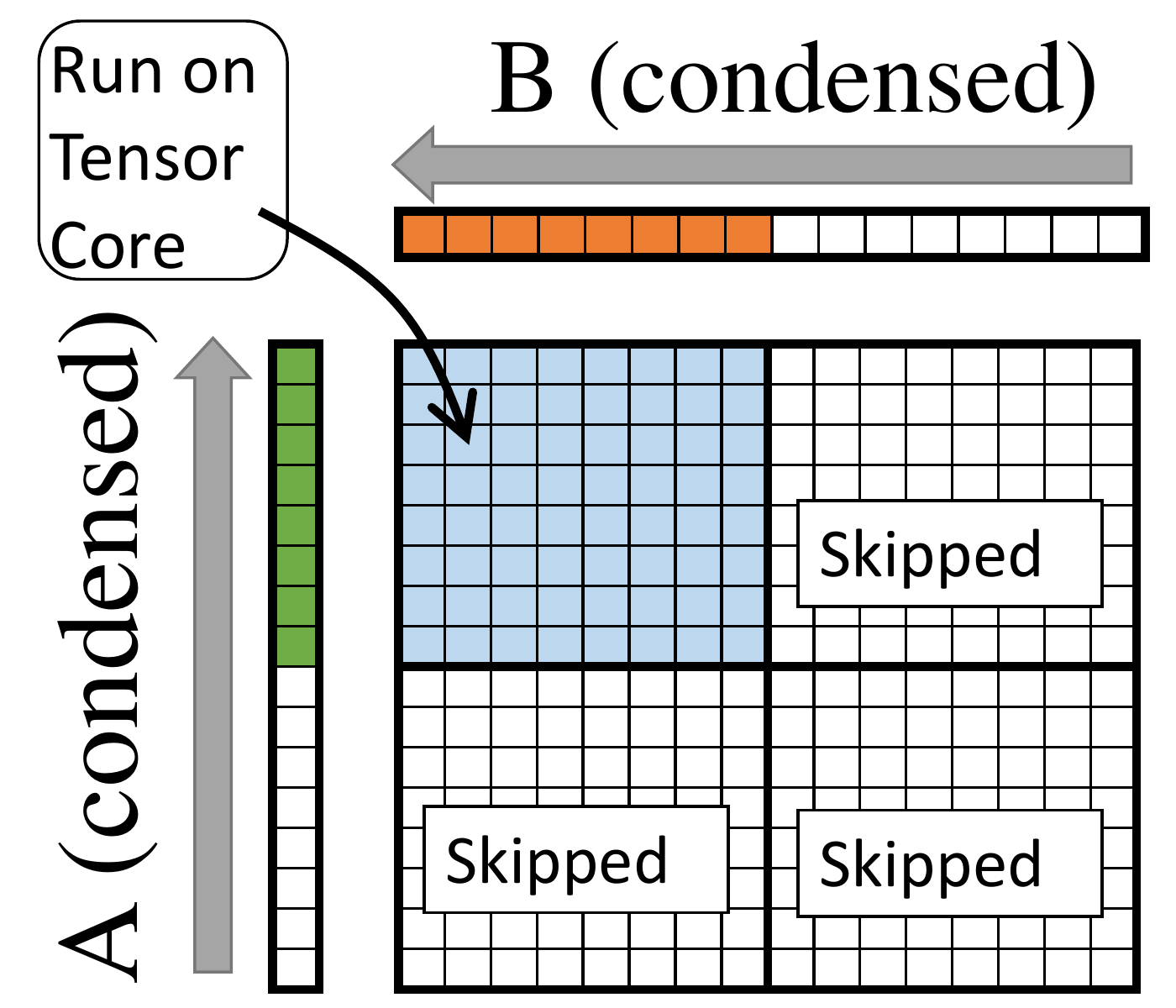}\label{fig:outer_prod_prob_2}}
\caption{\small(a) 8$\times$8$\times$1 matrix multiplication primitive in outer-product-based Tensor Core ; (b) Outer-product with two sparse inputs; (c) Leveraging dual-side sparsity with outer-product.}
\label{fig:outer_prod_}
\end{figure}

\subsection{SpGEMM in a Warp}
\label{sec:algo:warp}


\subsubsection{Problems of Inner-product Tensor Core}
In V100 architecture\cite{v100}, a warp controls two tensor cores simultaneously. \hl{Each tensor core can complete a $4\times4\times4$ dense matrix multiplication per cycle in a 4-stage pipeline\cite{raihan2019modeling}.} The basic computation unit in a dense tensor core is a parallel 4-element vector-vector dot product unit that multiplies and accumulates A matrix row and B matrix column, as Figure~\ref{fig:inner_prod_prob_1} shows. For single-side sparse matrix multiplication, the dot product requires selecting and accessing non-zero elements in matching positions in the dense vector. 
To solve the irregular addressing introduced in sparse models, sparse tensor core \cite{a100}\cite{mishra2021accelerating} uses a structural pruning scheme that conducts a 2-out-of-4 pruning in each partitioned sub-vector, which enforces a constant 50\% weight sparsity to balance the workload and to exploit parallelism in the dot-product unit, as shown in Figure~\ref{fig:inner_prod_prob_2}. However, this method is \textit{inefficient} for dual-side sparse matrix multiplication because activation sparsity is input-dependent and the number of non-zeros to be jointly matched is unpredictable, which results in difficulties to fulfill the dot product parallelism, as shown in Figure~\ref{fig:inner_prod_prob_3}. Although some prior ASIC designs \cite{sparten,scnn,extensor} propose dedicated hardware to solve this problem, their method either uses complex prefix sum hardware, costly shuffling register, or explicit barrier, which introduces considerable overheads and significant modifications to the Tensor Cores.

\subsubsection{Outer-product Tensor Core (OTC)}
Our design adopts an outer-product-based tensor core shown in Figure~\ref{fig:outer_prod_prob_0}.
We use an $8\times8\times1$ tile size as it has the same number of multipliers and adders (i.e., 64 in FP16) as the inner-product tensor core. 

OTC naturally avoids the inner-join process and can eliminate the irregular addressing by condensing two sparse inputs into two dense ones. 
As shown in Figure~\ref{fig:outer_prod_prob_2}, outer-product-based solutions can push all non-zeros in each column of matrix $\mathbb{A}$ to the upper-side and all non-zeros in each row of matrix $\mathbb{B}$ to the left, forming two dense vectors. Outer-product multiplication on the condensed inputs yields a condensed matrix multiplication. After condensing, non-zero elements are concentrated so that tensor cores take fewer instructions to complete a matrix multiplication and thus can achieve speedup over the original dense one. 

\begin{figure}[t]
\centering
\includegraphics[width=1\linewidth]{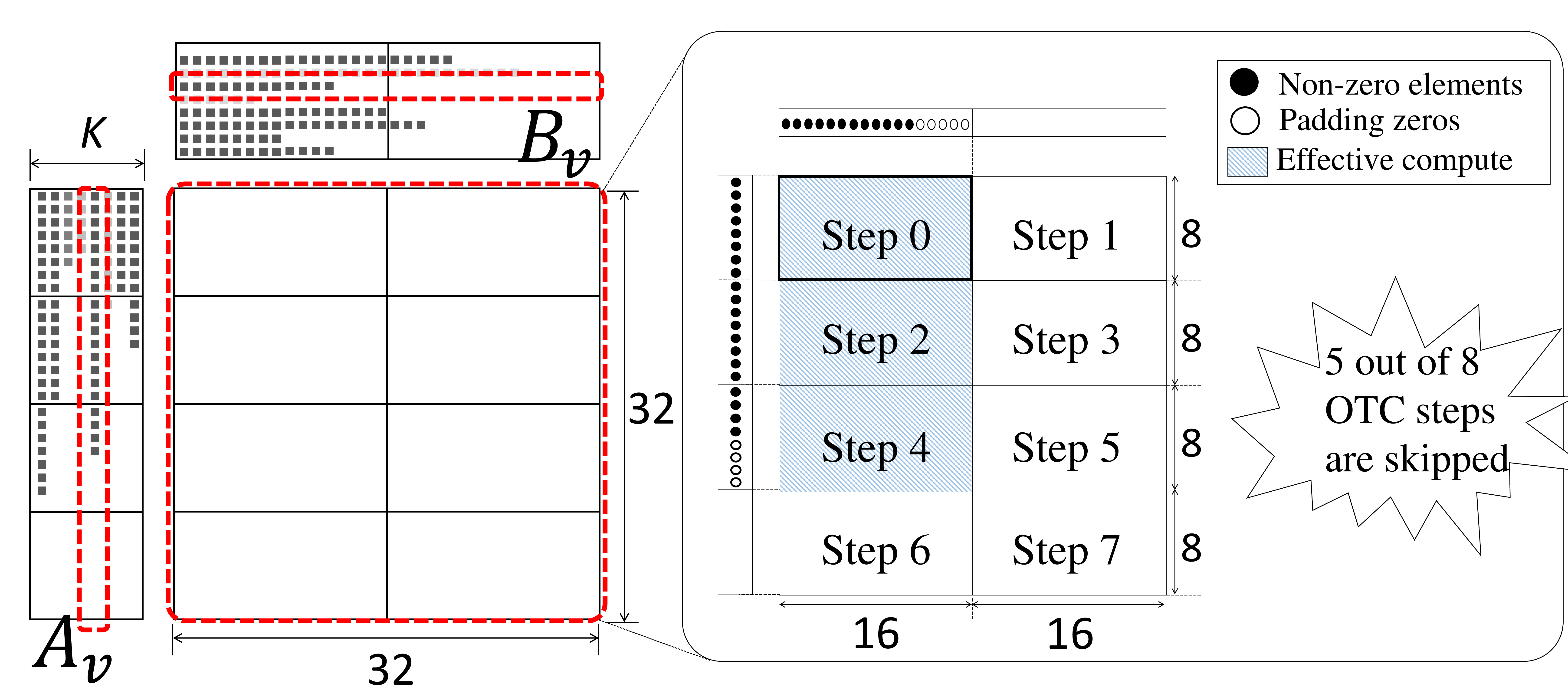}
\caption{SpGEMM in a warp.} 
\label{fig:spwarp}
\end{figure}

\begin{figure}[t]
\centering
\includegraphics[width=0.78\linewidth]{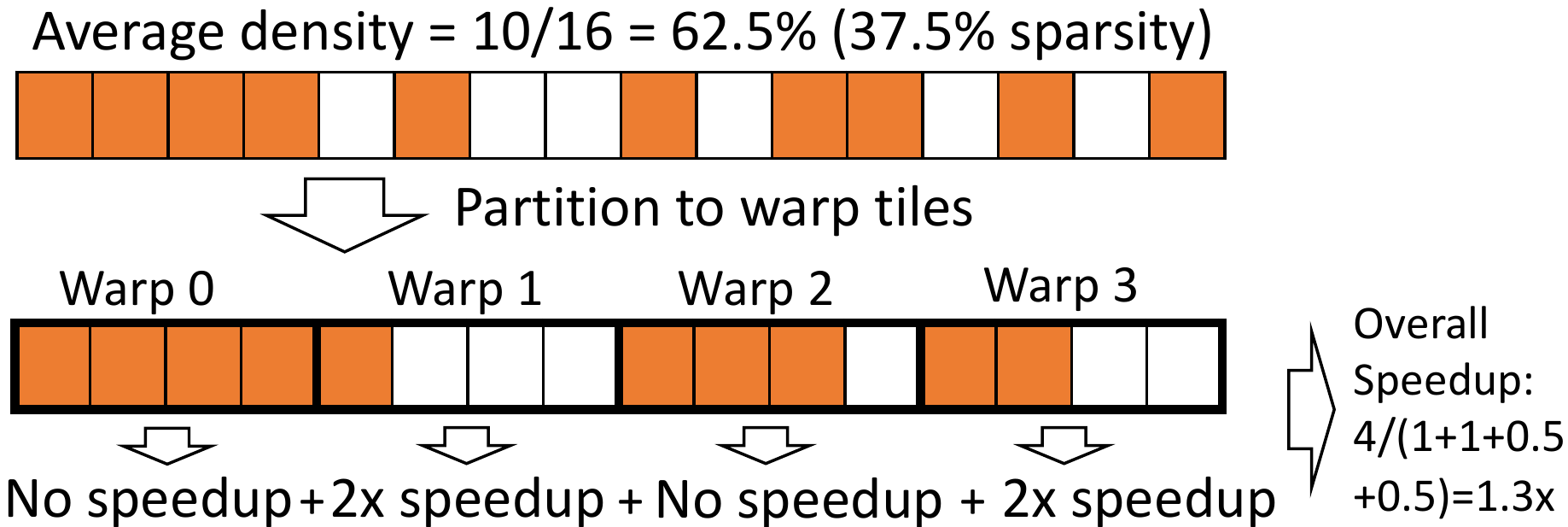}
\caption{\hl{Speedup on global matrix.}}
\label{fig:warp_speedup}
\end{figure}

\begin{figure}[t]
 \centering
  \includegraphics[width=0.95\linewidth]{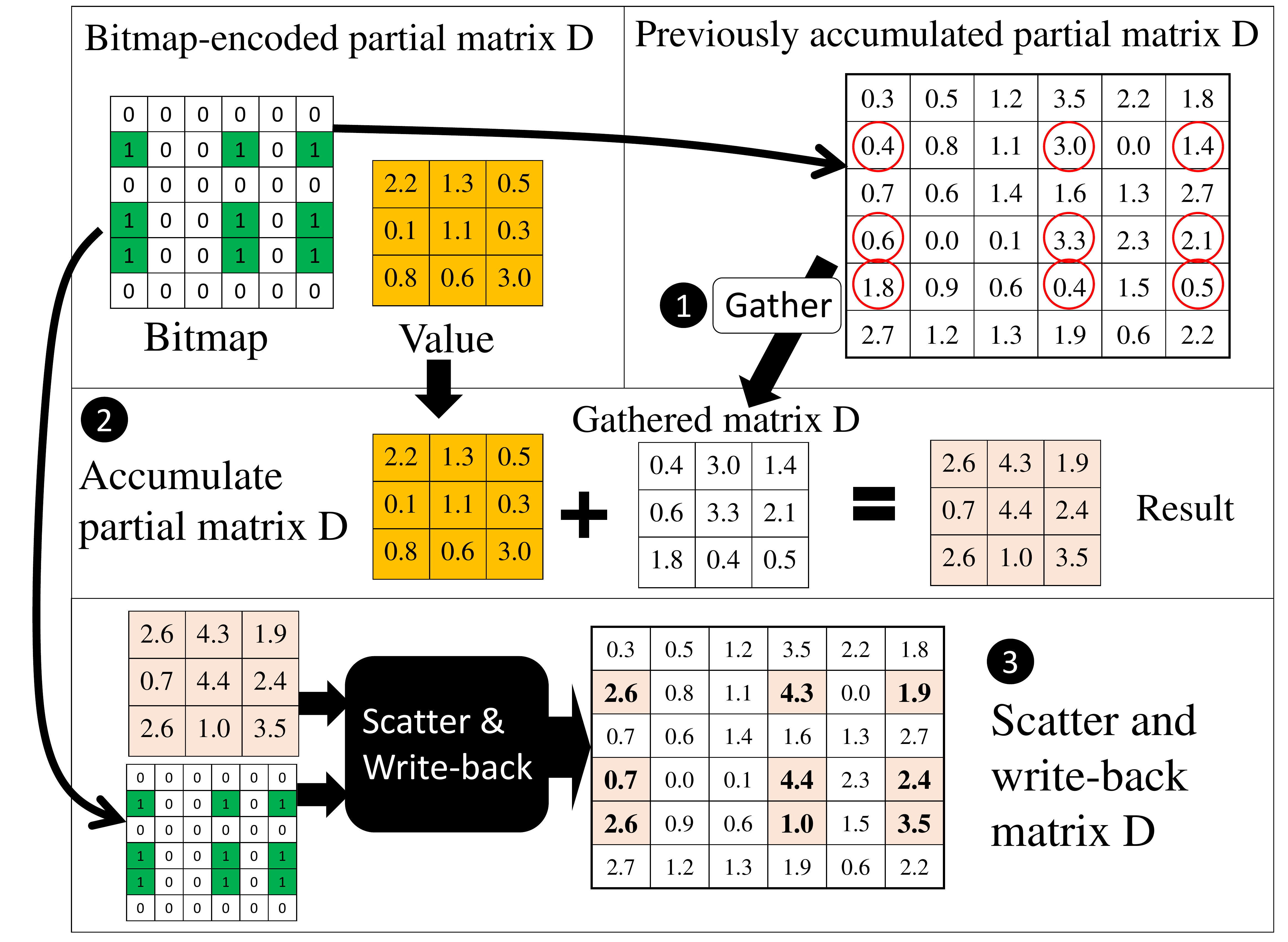}
  \caption{Gather-Scatter accumulation.}
  \label{fig:accumulator}
\end{figure}

\subsubsection{Warp-level Outer-product SpGEMM}

We propose an efficient warp-level SpGEMM algorithm with OTCs. Recall that there are two Tensor Cores working concurrently in a warp, each performing a $4\times4\times4$ matrix multiplication. 
In this section, we assume that the OTCs maintain an equivalent computing power, with two tensor cores together performing $8\times16\times1$ outer product. 
Section~\ref{sec:arch} will provide more architectural details.

Figure~\ref{fig:spwarp} shows an example where our warp-level SpGEMM achieves speedup on OTC with sparse inputs. It shows a $32\times32\times K$ warp tile computed in the outer-product manner. Since OTC computes an $8\times16\times1$ tile in a cycle, it takes 8 steps to complete the $32\times32\times1$ outer-product. For sparse inputs, $\mathbb{A}_v$ and $\mathbb{B}_v$ have fewer non-zeros elements in each column and row. Thus we can achieve speedup by skipping OTC steps. Figure~\ref{fig:spwarp}'s right-hand shows an example, where the column vector from $\mathbb{A}_v$ has 20 non-zeros in 32 elements, and row vector from $\mathbb{B}_v$ has 11 non-zeros in 32 elements. As such, 5 out of 8 OTC steps have pure zero elements and can be skipped, leading to a $\frac{8}{3}=2.67\times$ speedup in theory. The number of skipped OTC steps depends on the sparsity of the input vectors, which are $\langle 0\%,25\%,50\%,75\% \rangle$ on the $\mathbb{A}_v$ side and $\langle 0\%,50\%\rangle$ on the $\mathbb{B}_v$ side. Zeros are padded to the inputs to fulfill OTC's $8\times16$ tile dimension. 

\emph{Discussion:} \hl{Although the acceleration opportunity within a warp relies on an enumerable number of fixed sparsity ratios (e.g., $\langle 0\%,50\%\rangle$ for $\mathbb{B}_v$), our method at the global matrix level can go beyond this limitation. Figure~\ref{fig:warp_speedup} shows an example where a row of the global matrix has a 37.5\% sparsity, which should have no speedup on the assumption that we can only benefit from $\langle 0\%,50\%\rangle$ sparsity. However, we can still achieve an approximately 1.3$\times$ speedup after considering warp tiling at the global matrix level. Because the non-zeros are usually not evenly distributed across the global matrix, some warps such as warp 1 and 3 in Figure~\ref{fig:warp_speedup} can still enjoy the speedup provided by our SpGEMM.}

One design challenge in our warp-level SpGEMM algorithm is that the outer product requires all $M\times N$ elements of $\mathbb{D}$ to be stored in Tensor Core's local buffer so that it can be accumulated immediately. Therefore, the warp-tile size is majorly constrained by the Tensor Core's local buffer size. Hardware design and evaluation are described in Section~\ref{sec:arch} and Section~\ref{sec:exp}, respectively. 

\subsubsection{Merge}
\label{sec:algo:accumulate}

The \textit{merge} operation accumulates the partial matrices in multiple steps, as in Figure~\ref{fig:algo_overview_3}. In each step, we merge the newly generated partial matrix in bitmap encoding (e.g., $\mathbb{D}2_v$ and $\mathbb{D}2_b$) with the previously accumulated results (e.g., $\mathbb{E}1$). 
The pre-computed bitmap matrix (e.g., $\mathbb{D}1_b$) lets us easily derive the positions of non-zeros to be accumulated.  


Fig.~\ref{fig:accumulator} shows three steps in our \textit{merge} operation. First, we use the bitmap to \ding{202}\textit{gather} corresponding elements from the previous accumulated matrix. Then, the gathered elements are \ding{203}\textit{accumulated} to the values from \textit{multiply-value} output. 
We finally use \ding{204}\textit{scatter} function to restore non-zeros' positions by matching the 1's in bitmap and write to the result matrix. 

To map \textit{merge} on the OTC, we integrate the gather-accumulate-scatter procedure into Tensor Core's output matrix buffer with two optimizations. On one side, parallel accumulations are required to match OTC's multiply throughput. 
We design an efficient multiply-accumulate pipeline with 128-way parallel accumulators. On the other side, we design a light-weight operand collector to deal with irregular memory access in the gather-scatter function. Hardware design and evaluation are discussed in Section~\ref{sec:arch} and Section~\ref{sec:exp}.



\subsection{SpGEMM on the device}
\label{sec:algo:hier_tiling}



\begin{figure}[t]
\centering
\subfloat[][Tiled SpGEMM with one-level bitmap encoding.]{\includegraphics[width=1\linewidth]{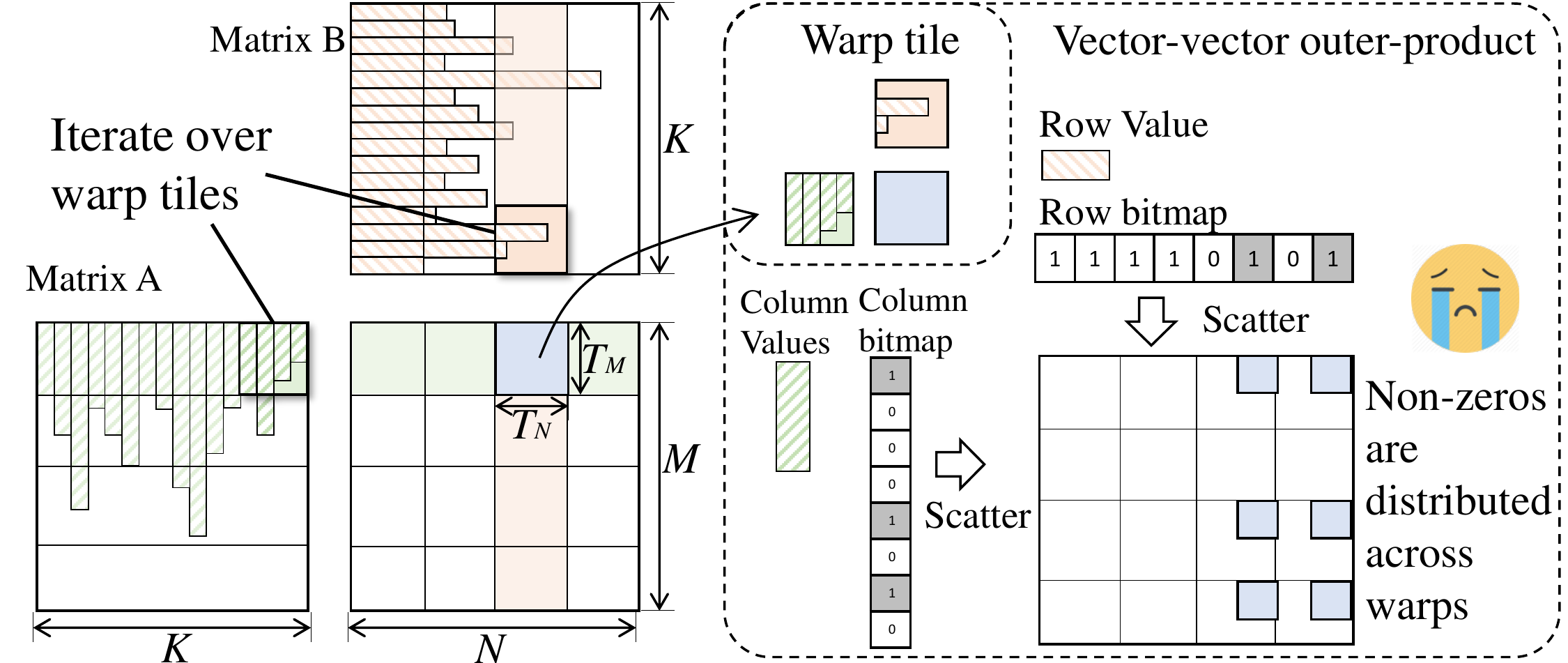}\label{fig:two_level_con_1}}\newline
\subfloat[][Tiled SpGEMM with two level bitmap encoding.]{\includegraphics[width=1\linewidth]{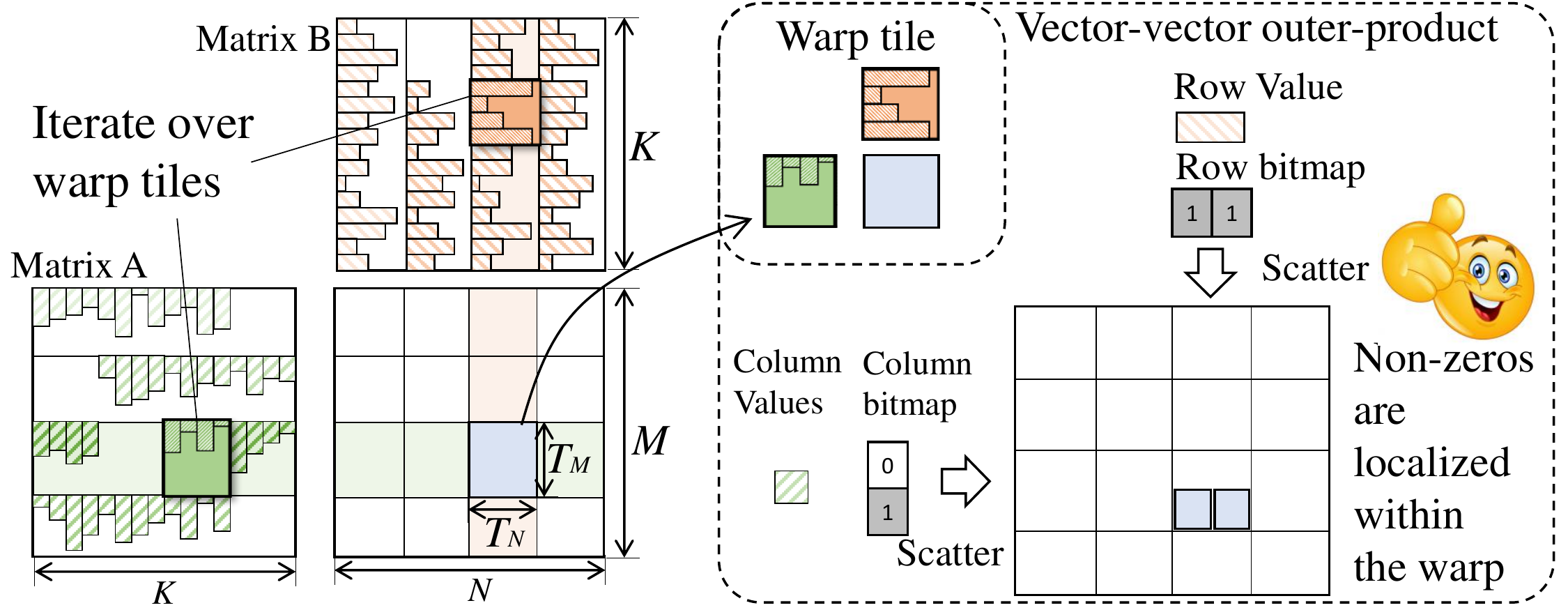}\label{fig:two_level_con_2}}
\caption{Data-locality aware sparse representation.}
\label{fig:locaity_condense} 
\vspace{-0.1in}
\end{figure}

\begin{figure}[t]
 \centering
  \includegraphics[width=0.8\linewidth]{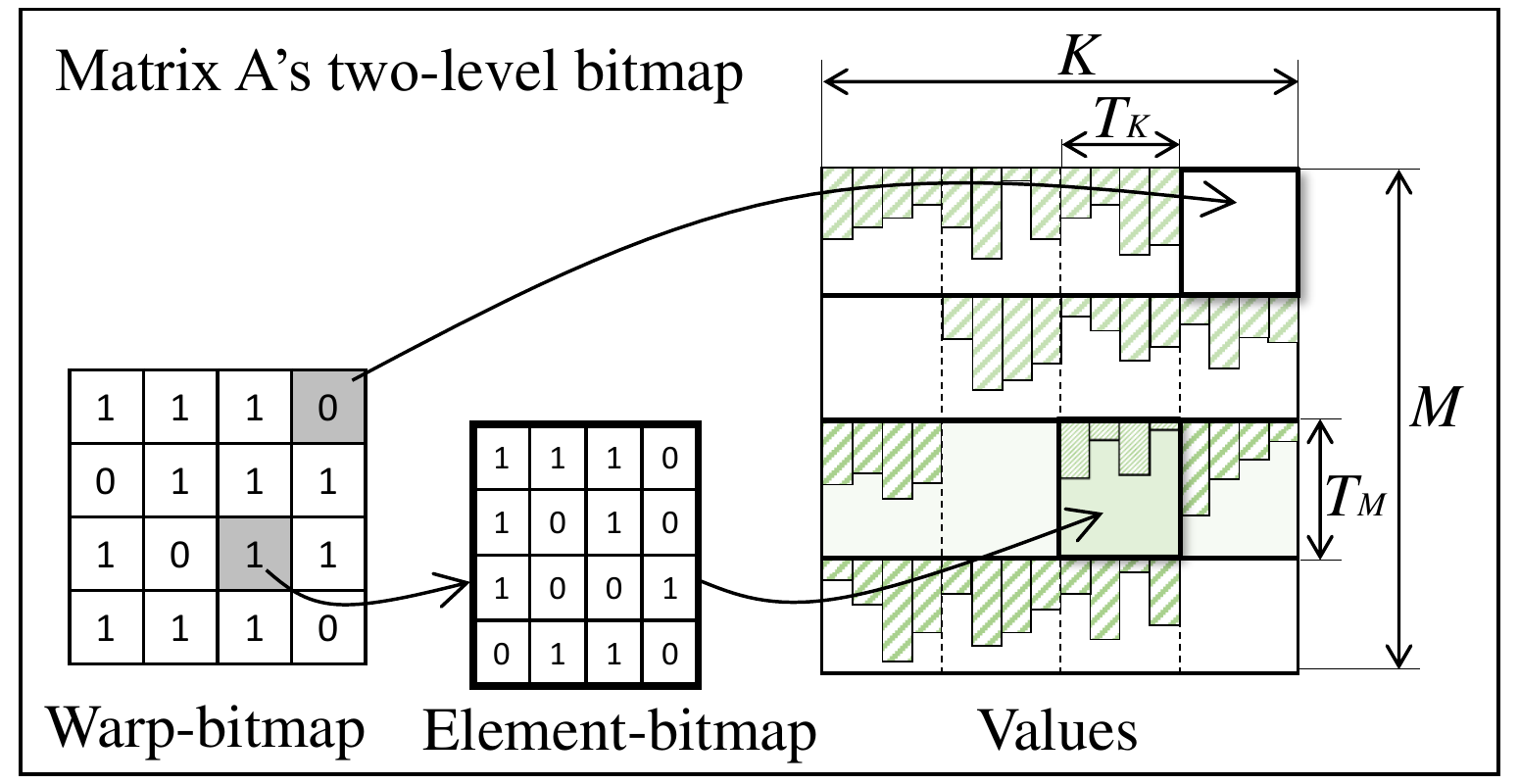}
  \caption{Two-level bitmap encoding format.}
  \vspace{-0.15in}
  \label{fig:two_level_con_3}
\end{figure}



The biggest challenge to map the SpGEMM on the entire device is that the outer product has poor output data reuse. When running a large matrix multiplication, outer products will produce a large amount of data in partial matrices. For SpGEMM, randomly distributed non-zero elements (e.g., $\mathbb{D}1$) in the partial matrix will yield a large addressing space and may often exceed a warp's local buffer size. This long-range addressing will result in frequent and fragmented global memory access, as the example shows in Figure~\ref{fig:two_level_con_1}.

To address the problem, we introduce a hierarchical bitmap encoding format that is aware of GPU's tiling scheme on SpGEMM. As Figure~\ref{fig:two_level_con_2} shows, we first partition SpGEMM into thread blocks, each of which computes an output block by iteratively loading blocks of A and B from input matrices. Each thread block computes a $32\times 32 \times 16$ matrix multiplication by running a warp-level SpGEMM aforementioned. 

To achieve this optimization, we propose a two-level bitmap encoding format. It contains three tuples, as shown in Figure~\ref{fig:two_level_con_3}. The first level bitmap encodes each partitioned matrix tile. Each `1' or `0' in this bitmap represents elemental zeros or non-zeros in the warp tile, and thus it is called element-bitmap. Since the sparse inputs' non-zero elements are located within the tile, the positions of output partial matrix non-zeros are also located within this tile, and thus can be fitted in Tensor Core's fastest local buffer and avoid external memory access. The second level bitmap is called warp-bitmap, which uses a `1' or `0' to represent the entire tile, where `0's means the tile is empty and `1's is not. The warp with a `0' warp-bit can be skipped because either of the two input tiles is pure zeros.

\begin{figure}[t]
\centering
\subfloat[][Inner product friendly im2col.]{\includegraphics[width=0.95\linewidth]{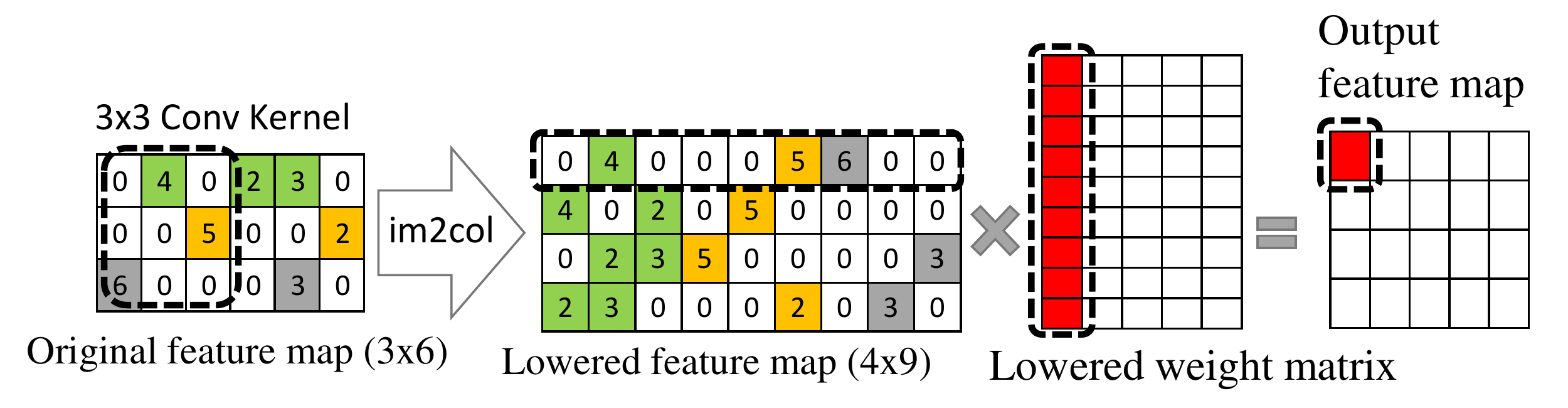}\label{fig:inner_friend_im2col}} \\ 
\subfloat[][Outer product friendly im2col.]{
\includegraphics[width=0.95\linewidth]{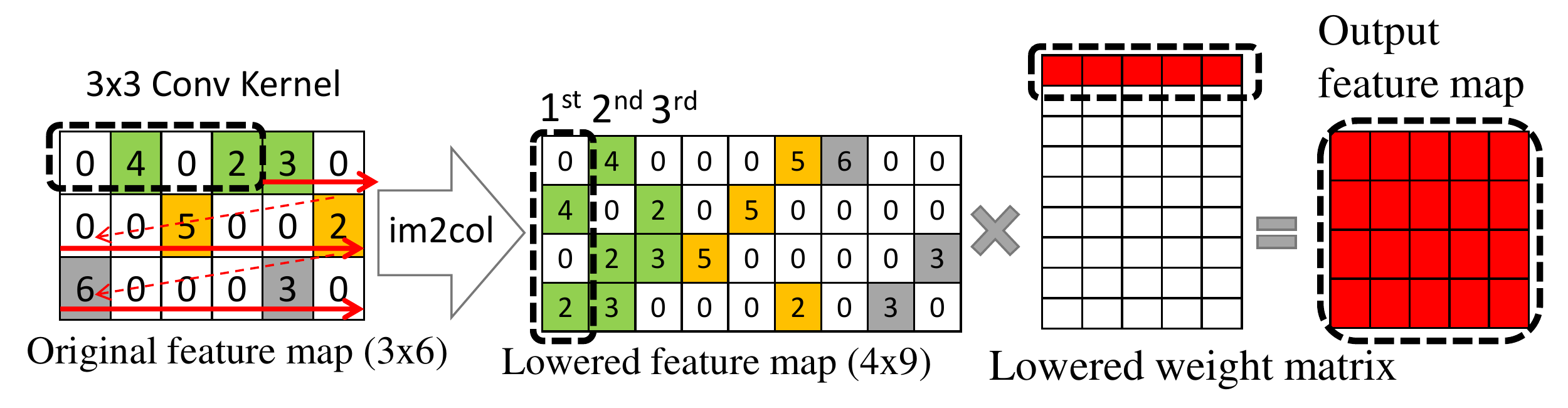}\label{fig:outer_friend_im2col}
}
\caption{Outer product friendly im2col on dense matrix.} 
\label{fig:outer_im2col}
\end{figure}

\section{Dual-side Sparse Convolution}
\label{sec:conv}


GPU usually accelerates dense convolution by transforming it into a GEMM operator through im2col function. To leverage the proposed SpGEMM algorithm, we propose a novel sparse im2col method for dual-side sparse convolution. 

Im2col mainly rearranges the data organization of input feature maps as an input of GEMM. Thus, improperly designed im2col may harm input data re-use for matrix multiplication. 
We propose an outer-product friendly im2col method for generic outer-product. To avoid the space and time overhead by doing im2col explicitly, we present our \textit{implicit} sparse im2col algorithm that supports efficient data rearranging in registers.

\subsection{Outer-product friendly im2col}
\label{sec:algo:outer_friend}

Figure~\ref{fig:inner_friend_im2col} shows an example of im2col on a $3\times 6$ feature map with a $3\times 3$ convolution kernel. It rearranges all elements in the $3\times 3$ sliding window into a row in the lowered feature map. And the output matrix of im2col is generated by shifting the sliding window by one element per step with multiple rounds. This process is friendly for inner-products because one row in each step matches inner-product's multiply-and-accumulated computation. On the contrary, outer-product requires a column of data in each step, which im2col cannot utilize the fully lowered feature map. 

In contrast, the proposed outer-product-friendly im2col generates a column of data in the lowered feature map at a time in Figure~\ref{fig:outer_friend_im2col}. For example, the first three columns come from the first row in the feature map. And these three columns share four data from each other. If sliding a 1$\times$4 window scanning over the feature map in a zig-zag way, we will get a column-major lowered feature map as the input of GEMM. In generic cases, the number of values in a column is decided by feature map and convolution kernel size, which is calculated by $B = (R-K+S)/S$, where $\langle R, K, S \rangle$ stands for row size of feature map, convolution kernel size, and stride size, respectively. This transformation is equivalent to permuting the loop nest of accessing the lowered feature map by structuring the inner-most loop as the outer-most. 

\subsection{Bitmap-based sparse im2col}

\begin{figure}[t]
\centering
\subfloat[][Bitmap im2col's input and output.]{\includegraphics[width=0.78\linewidth]{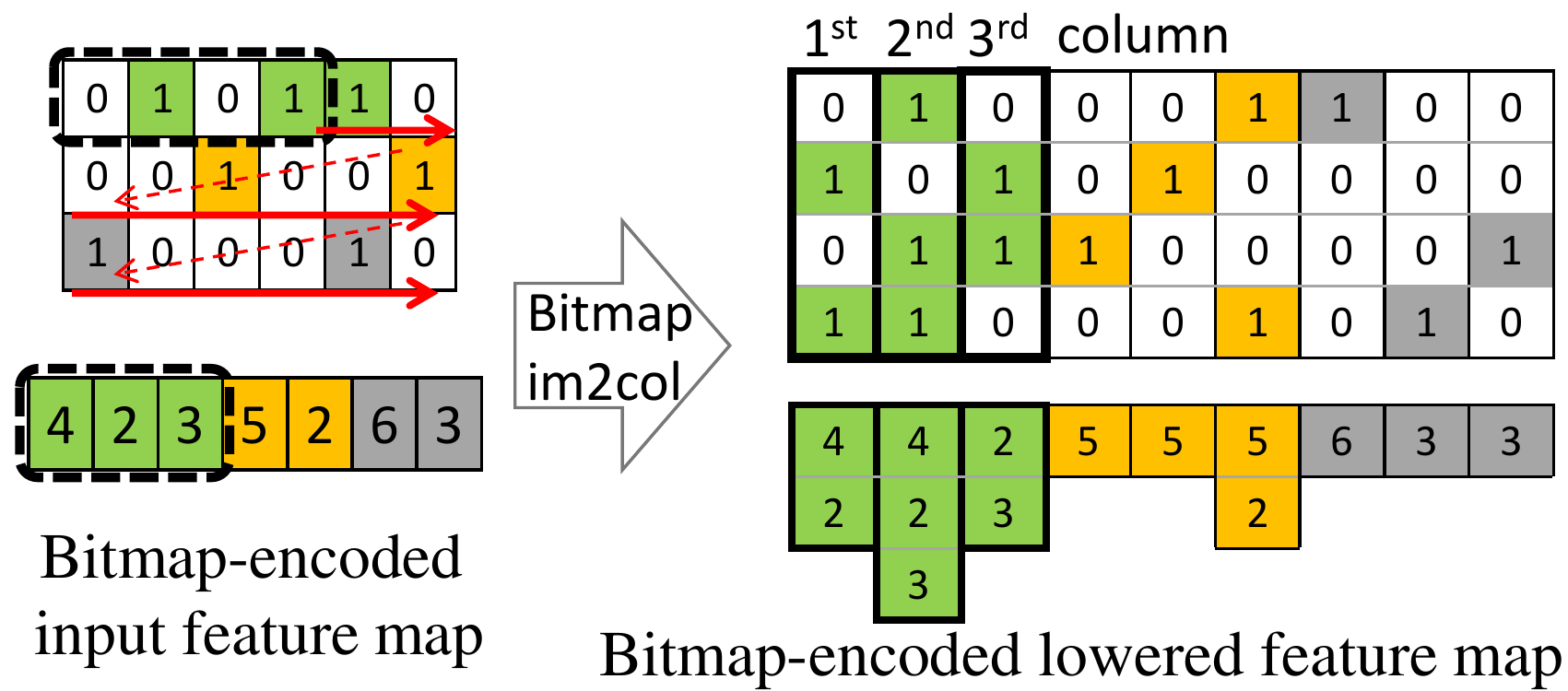}\label{fig:bitmap_im2col_a}}
\\ 
\subfloat[][The proposed sparse im2col on bitmap. \hl{This bitmap format comprises three fields. The bitmap field uses 0/1 to represent zero and non-zero positions. The value field contains non-zeros. The row offset is specifically introduced for the convenience of sparse im2col.}]{
\includegraphics[width=0.95\linewidth]{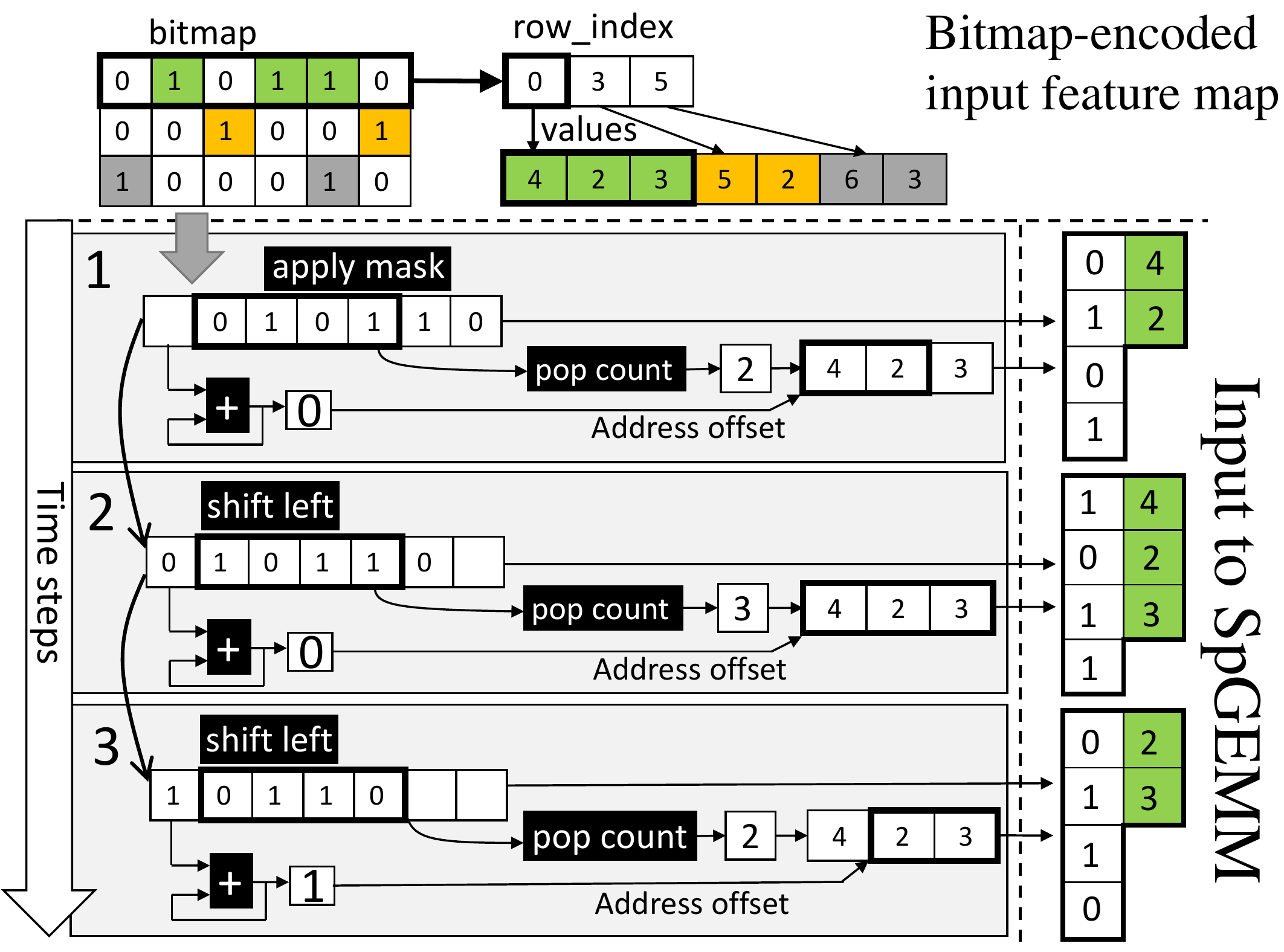}\label{fig:bitmap_im2col_b}
}
\caption{The im2col on a bitmap-encoded feature map.}
\vspace{-0.1in}
\label{fig:bitmap_im2col}
\end{figure}

Similar to dense \emph{implicit im2col}, our \emph{sparse implicit im2col} keeps the bitmap-encoded sparse feature maps in global memory and re-arranges data layout in registers. 
Matrix B is simply bitmap-encoding of the flattened sparse weight matrix.

Bitmap is efficient for sparse im2col because it inherits the structural information from a dense matrix. 
So bitmap-based encoding format can first conduct im2col on bitmap using a method similar to dense im2col and generate a lowered bitmap. Then, we use the bitmap as a mask to fetch the corresponding non-zero values. Figure~\ref{fig:bitmap_im2col} shows a detailed flow of our approach with an example. We use a $3\times 6$ feature map convoluted by a $3\times 3$ kernel, and get a $4\times 9$ lowered feature map in Figure~\ref{fig:bitmap_im2col_a}. Due to the outer-product-friendly im2col method, the first three columns of data, which we highlight, are generated sequentially in this example.


\begin{itemize}[leftmargin=*]
    \item[S0] We first encode the original feature map in bitmap format. 
    \item[S1] We take the first bitmap row from bitmap encoding and its corresponding non-zero values.
    \item[S2] For the first column, we \textit{apply a mask} on the bitmap row and output the bits falling in the mask as the first column bitmap for the lowered feature map. For the subsequent columns, we \textit{shift left} the bitmap row, which leads to shifted-out bit.
    \item[S3] We \textit{accumulate} the shifted-out bit and use the accumulated result as the address offset to access the non-zero values. 
    \item[S4] We use \textit{population count} to count the number of non-zeros in the mask. We find corresponding non-zero data in the value vector and output its value, address offset and length.
\end{itemize}

\hl{Our approach is efficient for two reasons. First, all required operations are low cost and can be conducted in register files. Second, the result is already in the condensed format and can be directly fed to outer-product SpGEMM via register reads.}

\section{Outer Product Sparse Tensor Core}
\label{sec:arch}

\begin{figure*}[t!]
\centering
\includegraphics[width=0.95\linewidth]{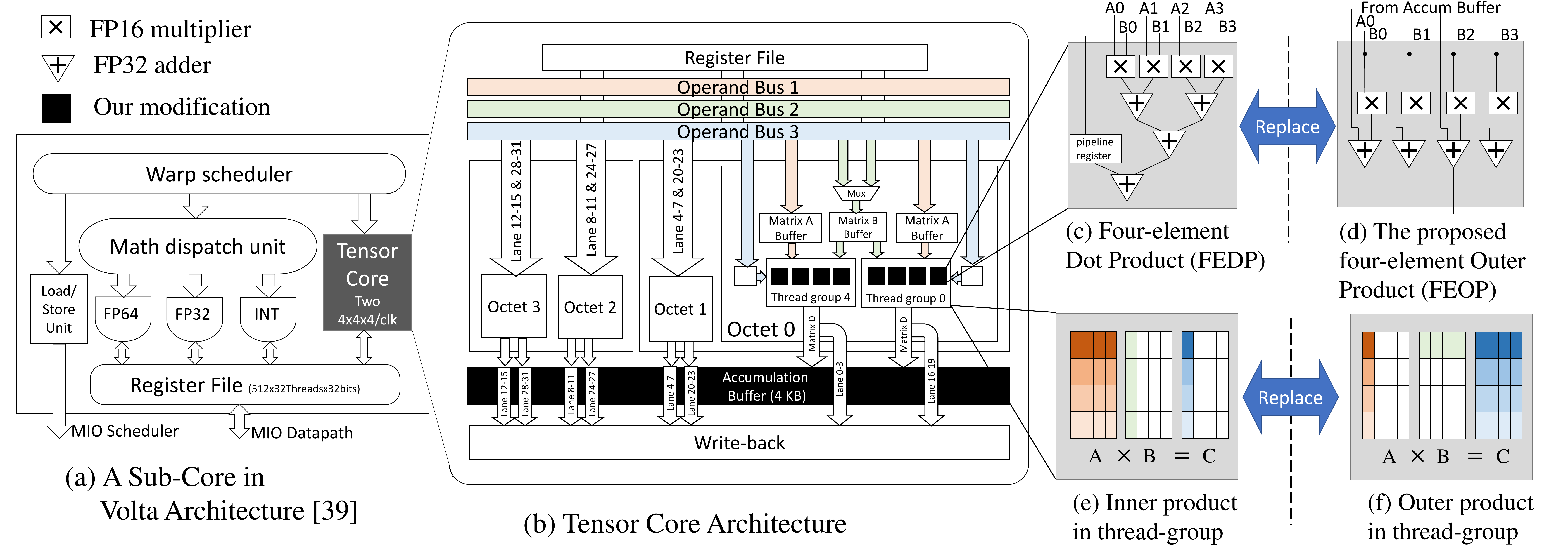}
\caption{\hl{Our modification to the Tensor Core includes the accumulation buffer in (b) and the replacement of Four-Element Dot Product Unit (c) with Four-Element Outer Product (d). Since there are four FEDPs in each thread group, the thread-group level inner-product (e) is replaced with outer-product (f).}}
\vspace{-0.2in}
\label{fig:tensorcore_arch}
\end{figure*}

In this section, we introduce the micro-architecture extensions to support our bitmap-based SpGEMM and SpCONV. 

\subsection{Outer-product Tensor Core (OTC)}

We modify Tensor Core hardware from inner-product to outer-product for \textit{dense matrix multiplication} because it is a pre-requisite for our SpGEMM algorithm.

\subsubsection{Hardware modification}
Tensor Core, a specialized matrix multiplication hardware, is integrated into NVIDIA's GPGPU since Volta architecture\cite{v100} to accelerate machine learning tasks. Figure~\ref{fig:tensorcore_arch}a shows an overview of a Sub-Core \cite{volta} in a Volta's streaming processor (SM). Each Sub-Core contains a bunch of math function units and two Tensor Cores. 
\hl{Each Tensor Core completes a $4\times4\times4$ dense matrix multiplication in a cycle in a 4-stage pipeline\cite{raihan2019modeling}.} In the V100 GPU, a total number of 640 Tensor Cores are distributed across 80 SMs, with 2 Tensor Cores per Sub-Core and 4 Sub-Cores per SM, providing a peak performance of 125 TFLOPS at 1530 MHz clock frequency.

Figure~\ref{fig:tensorcore_arch}b shows a detailed architecture of the two Tensor Cores in a Sub-Core. Each Tensor Core contains 16 inner-product units. Each inner product unit performs a four-element dot-product (FEDP), yielding a total computing power of 64 multiply-accumulate per cycle in a single Tensor Core. Figure~\ref{fig:tensorcore_arch}c details a FEDP structure, which multiplies and accumulates two four-element vectors from A and B in parallel. The 16 FEDPs are grouped into two `Octets,' eight to each Octet. One Octet is further split into two thread groups. Each thread group contains four FEDPs and computes four elements in the $4\times4$ output matrix, as is shown in Figure~\ref{fig:tensorcore_arch}e. 

We modify the above-mentioned inner-product Tensor Core to fit for \textit{dense outer-product}'s computation. Figure~\ref{fig:tensorcore_arch}d depicts our changes to the FEDP hardware. Our four-element outer product (FEOP) multiplies one element from A with four elements from B in parallel and accumulates partial results with the adders. As such, four FEOPs in a thread group collectively perform a $4\times4$ outer-product, as in Figure~\ref{fig:tensorcore_arch}f.

\subsubsection{ISA extensions for dense outer product}

\begin{figure}[t]
\centering
\subfloat[][Original inner-product WMMA.  ]{\includegraphics[width=0.48\linewidth]{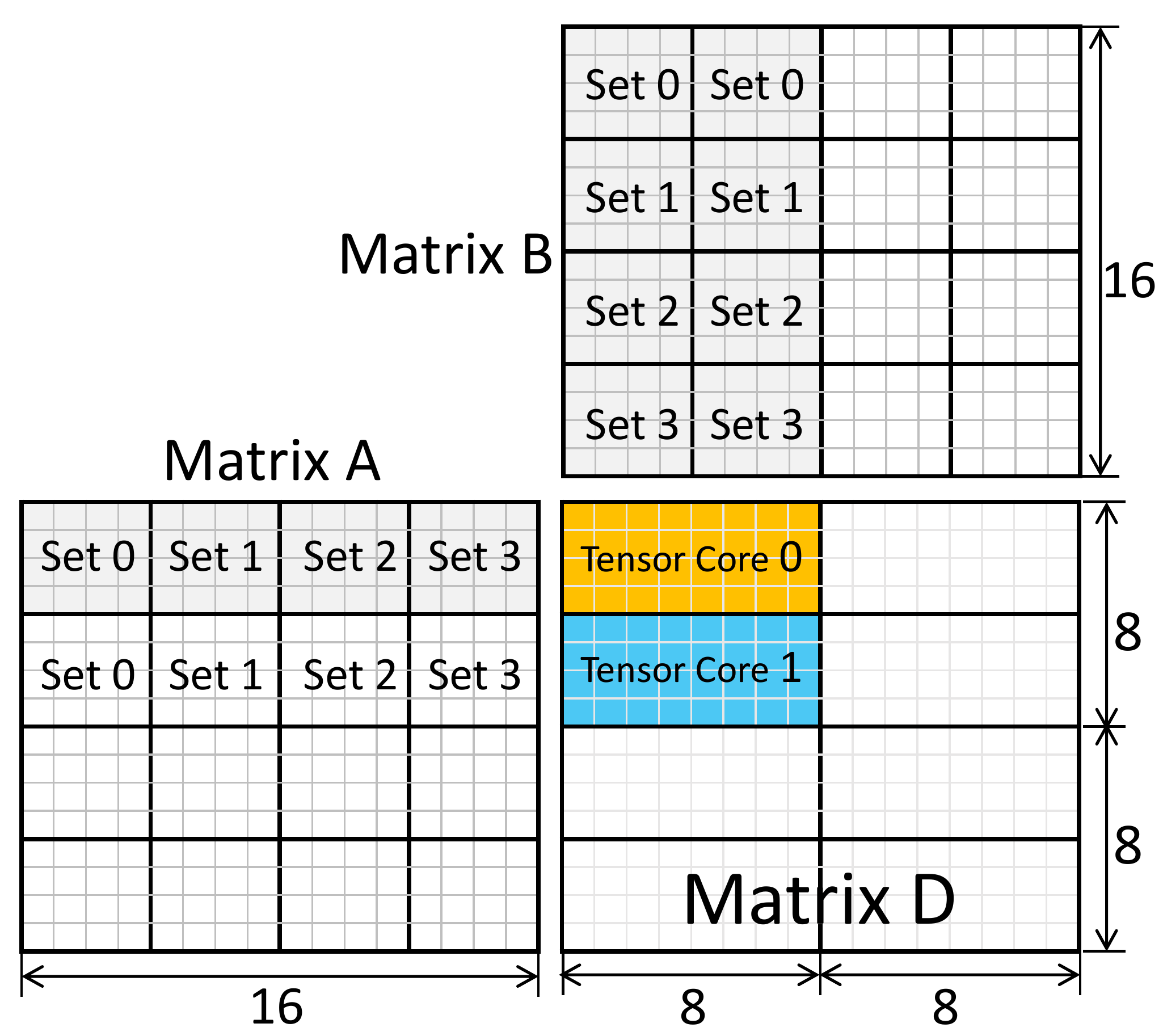}\label{fig:wmma_native}}
\subfloat[][Proposed outer-product WMMA.]{
\includegraphics[width=0.48\linewidth]{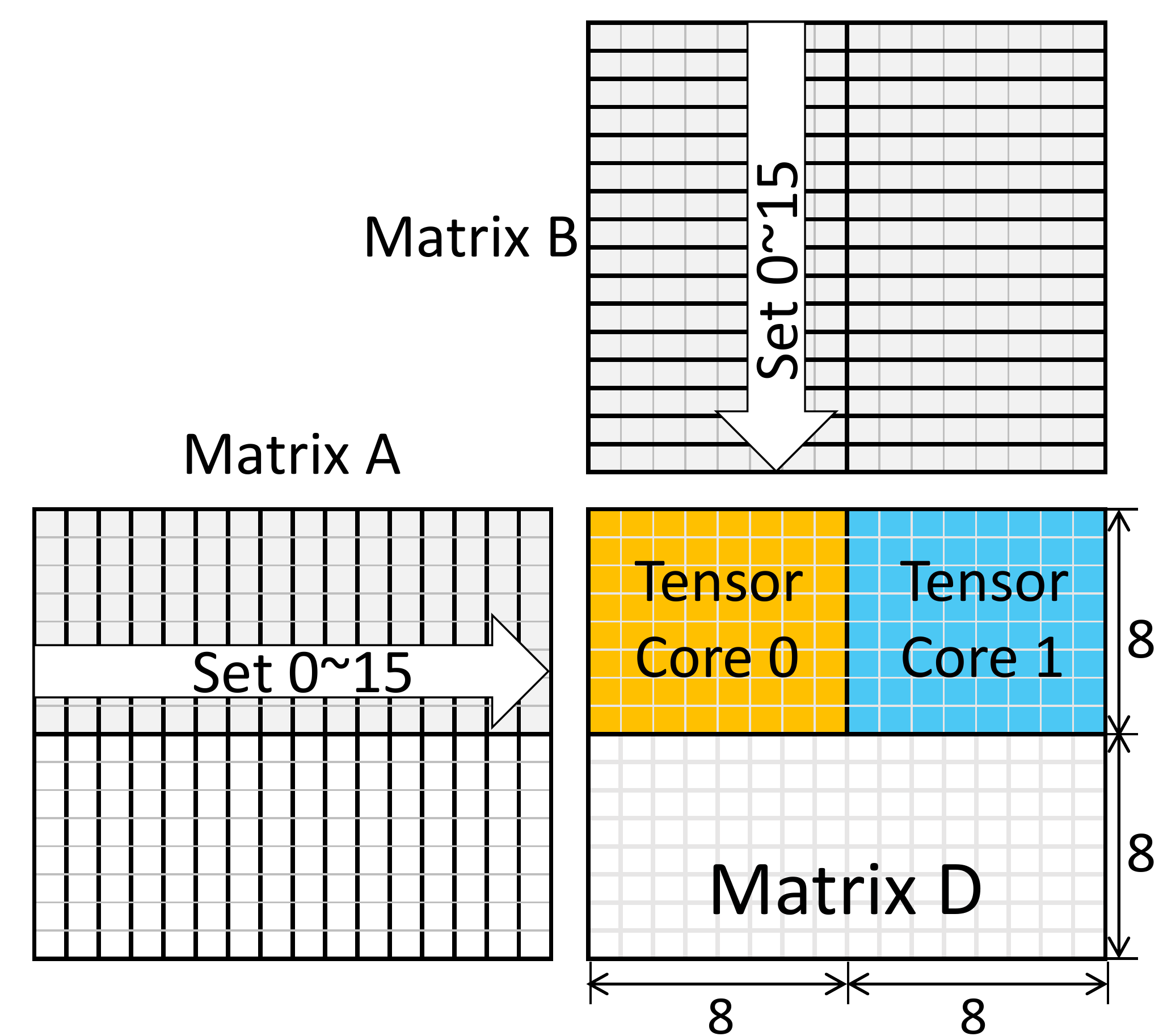}\label{fig:wmma_our}
}
\caption{\small The original WMMA operation computes a 16$\times$16$\times$16 matrix multiplication in a warp tile. We define OWMMA (outer-product WMMA) operation that computes the dense warp-tile with 16 sets of OHMMA.8161 instructions.}
\label{fig:wmma_outer_prod}
\vspace{-0.2in}
\end{figure}

\hl{Recall that each tensor core performs a $4\times4\times4$ dense matrix multiplication.} Two tensor cores work cooperatively and form a machine-level HMMA.884 instruction to compute an $8\times8\times 4$ output block by taking an $8\times4$ tile of A and a $4\times8$ tile of B, as shown in Figure~\ref{fig:wmma_native}. Four sets of HMMA instructions are used to compute an $8\times8\times 16$ tile. At the warp level, CUDA exposes a WMMA API that computes a larger $16\times16\times16$ matrix operation with these HMMA instructions in 32 cycles.

Fig.~\ref{fig:wmma_our} depicts our outer-product tensor core (OTC) interface. Each OTC conducts an $8\times 8\times 1$ dense vector-vector outer product, which has the same 64 FP16 multipliers as the original $4\times4\times4$ Tensor Core. Two OTCs form a $8\times 16\times 1$ Outer-product HMMA (OHMMA.8161) instruction that takes a $8\times1$ tile of $\mathbb{A}_v$ and a $1\times16$ tile of $\mathbb{B}_v$ as input. We define 16 sets of OHMMA instructions to complete the full-warp $16\times 16\times 16$ matrix multiplication (OWMMA). In total, an OTC also takes 32 cycles to finish an OWMMA operation. 

The bitmap outer-product is also an essential step in our bitmap-based SpGEMM. To accelerate bitmap operation, we execute \textit{multiply-bitmap} on the OTC in Tensor Core. We define Binary OHMMA instruction (BOHMMA), which conducts outer product on 1-bit inputs. Since Volta architecture\cite{volta}, Tensor Core has already started to support binary operations that process 16$\times$ larger matrix tile than the FP16 operations\cite{choquette2020nvidia}. We inherit the binary operator design from the native Tensor Core and extend the binary OHMMA (BOHMMA) instruction to support $32\times32\times1$ binary outer product. The \textit{multiply-bitmap} achieves low cost because BOHMMA is 16$\times$ faster than HMMA on FP16 outer product. OHMMA and BOHMMA instructions are defined in Figure~\ref{fig:ohmma}. 

\begin{figure}[h]
\begin{minted}
[
    frame=single,
    fontsize=\scriptsize
]
{c}
HMMA.OHMMA.8161.F32.F32 {R9, R10, R11, R12}, 
              {R1, R2}, {R3, R4}, {R5, R6, R7, R8};
HMMA.BOHMMA.32321.B32.B32 R3, R1, R2;
\end{minted}
\caption{\small Extended machine-level OHMMA/BOHMMA instructions.}
\label{fig:ohmma}
\end{figure}


\subsection{Dual-side Sparse Tensor Core}

\begin{figure}[t]
\centering
\includegraphics[width=1\linewidth]{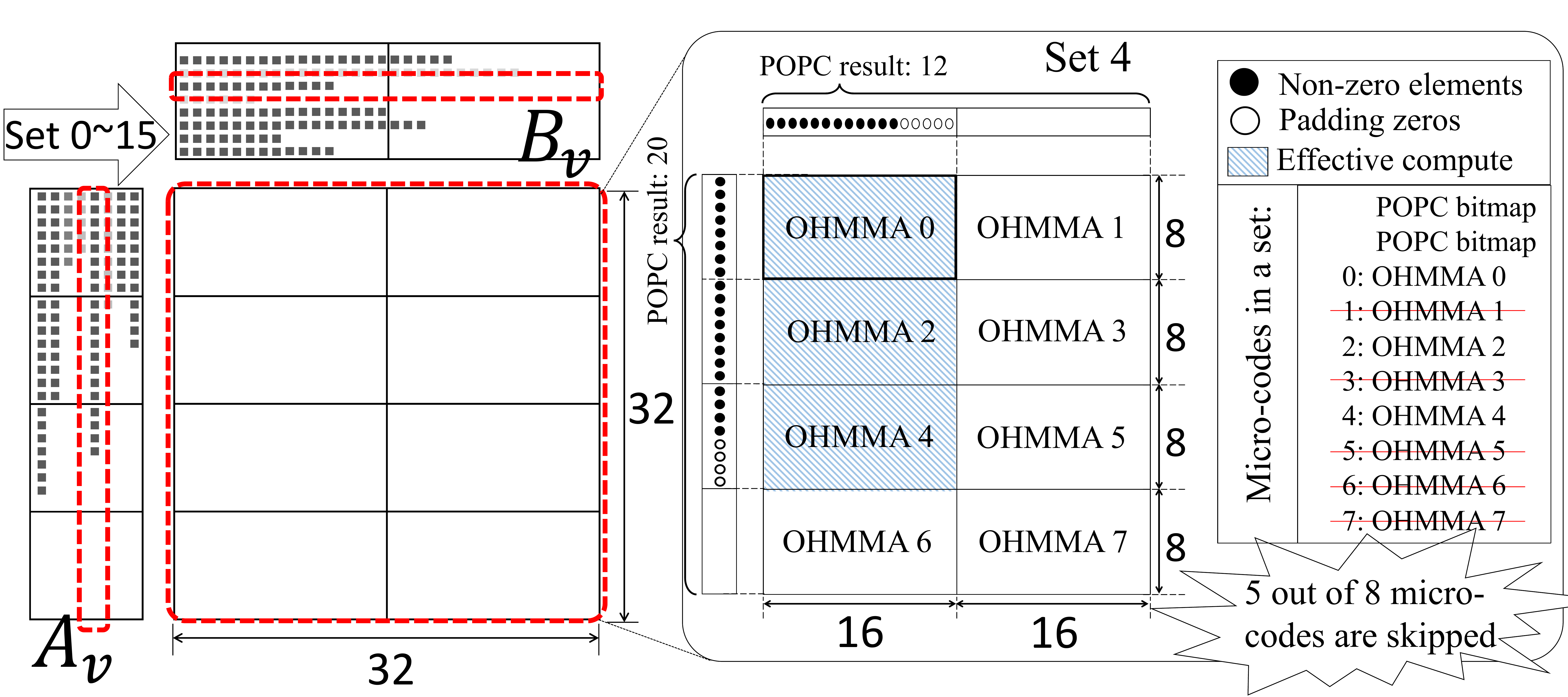}
\caption{\small \hl{The proposed SpWMMA includes 8 OHMMA instructions in dense mode, which can be skipped during the sparse mode.}} 
\vspace{-0.1in}
\label{fig:spwmma_exe}
\end{figure}

We propose two adaptations to achieve speedup on dual-side sparsity with the above hardware and instruction extensions support. On the software side, we define SpWMMA, warp-level dual-side sparse matrix multiplication API that exploits sparsity in matrix A and B by dynamically skipping OHMMA instructions. On the hardware side, we propose the accumulation buffer to gather partial results from outer-product units. 

\subsubsection{Warp-level interface} We define a SpWMMA API that works on a warp-level matrix tile in Figure~\ref{fig:spwmma_api}. A SpWMMA breaks down to 16 sets, and each set includes a $32\times32\times1$ outer product in Figure~\ref{fig:wmma_outer_prod}. Since the machine-level OHMMA instruction computes an $8\times16\times1$ outer product within a warp. And each SpWMMA API call is complied to 8 OHMMA instructions, as shown in Figure~\ref{fig:spwmma_breakdown}. 


For sparse inputs, $\mathbb{A}_v$ and $\mathbb{B}_v$ have fewer non-zeros elements and thus achieve speedup by skipping OHMMA instructions with predication operations. Predication operation is widely used in GPGPU to skip instruction executions. \hl{We utilize population count instructions (POPC, commonly supported in GPGPU to count the number of ``1" in binary numbers) to set predication bits of OHMMA instructions. The number of ``1" in $\mathbb{A}_v$'s and $\mathbb{B}_v$'s bitmaps identify the number of element-wise multiplication in each row/column of the condensed sparse matrix multiplication. By counting ``1" bits in the bitmap with POPC, we can determine which OHMMA instructions should be enabled. The right side of Figure~\ref{fig:spwmma_exe} shows an example of Set 4's computation with POPC and OHMMA instructions. We count $\mathbb{A}_v$'s and $\mathbb{B}_v$'s bitmaps, indicating that the sparse multiplication takes 12/20 multiplications in each row/column. In our design, each OHHMA instruction covers 8$\times$16 condensed sparse outer product multiplication. We should enable OHHMA0/2/4 by setting predication bits and skip OHHMA1/3/4/5/6/7 for Set 4.}


 \begin{figure}[h]
\begin{minted}
[
    frame=single,
    fontsize=\scriptsize
]
{c}
SPWMMA.MMA.SYNC.A_LAYOUT.B_LAYOUT.M32N32K1.set.f32.f32 
{%RD0~%RD7}, {%RB0~%RB7}, {%RA0~%RA7}, {%RC0~%RC7};
\end{minted}
\caption{\small Our SpWMMA API.}
\label{fig:spwmma_api}
\begin{minted}
[
    frame=single,
    fontsize=\scriptsize
]
{c}
HMMA.BOHMMA.32321.B32.B32 R3, R1, R2;
// ...
@p0 HMMA.OHMMA.8161.F32.F32 {R8, R9, R10, R11}, 
    {R4, R5}, {R6, R7}, {R8, R9, R10, R11};
@p1 HMMA.OHMMA.8161.F32.F32 {R16, R17, R18, R19},
    {R12, R13}, {R14, R15}, {R16, R17, R18, R19};
// ...
@p7 HMMA.OHMMA.8161.F32.F32 {R119, R120, R121, R122}, 
    {R115, R116}, {R117, R118},{R119, R120, R121, R122};
\end{minted}
\caption{\small SpWMMA API complied to machine-level instructions.}
\vspace{-0.4in}
\label{fig:spwmma_breakdown}
\end{figure}  

\subsubsection{Accumulation buffer}

\begin{figure}[ht]
\centering
\subfloat[][Dense mode.  ]{\includegraphics[width=0.44\linewidth]{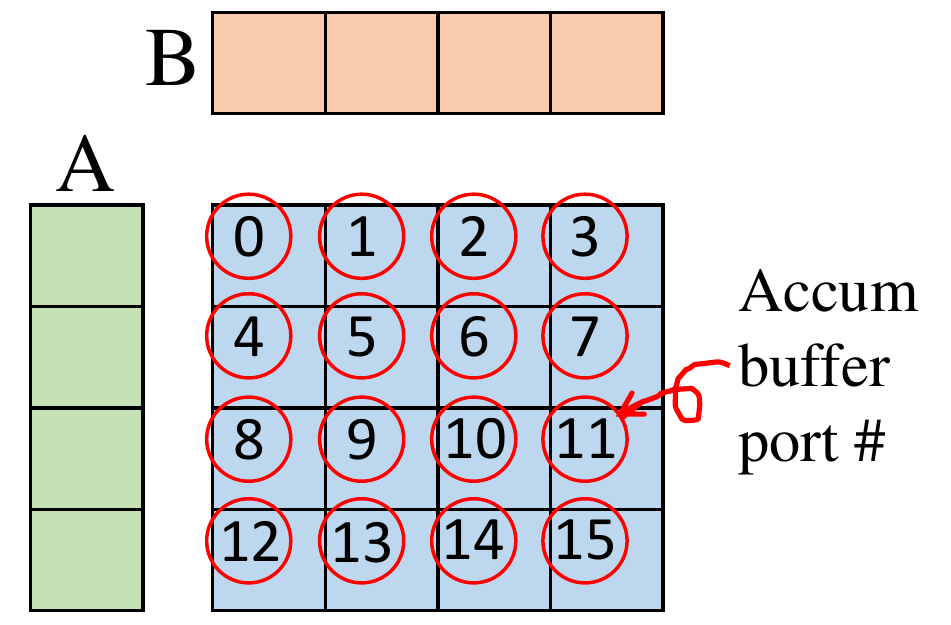}\label{fig:accum_dense}}
\subfloat[][Sparse mode.]{
\includegraphics[width=0.45\linewidth]{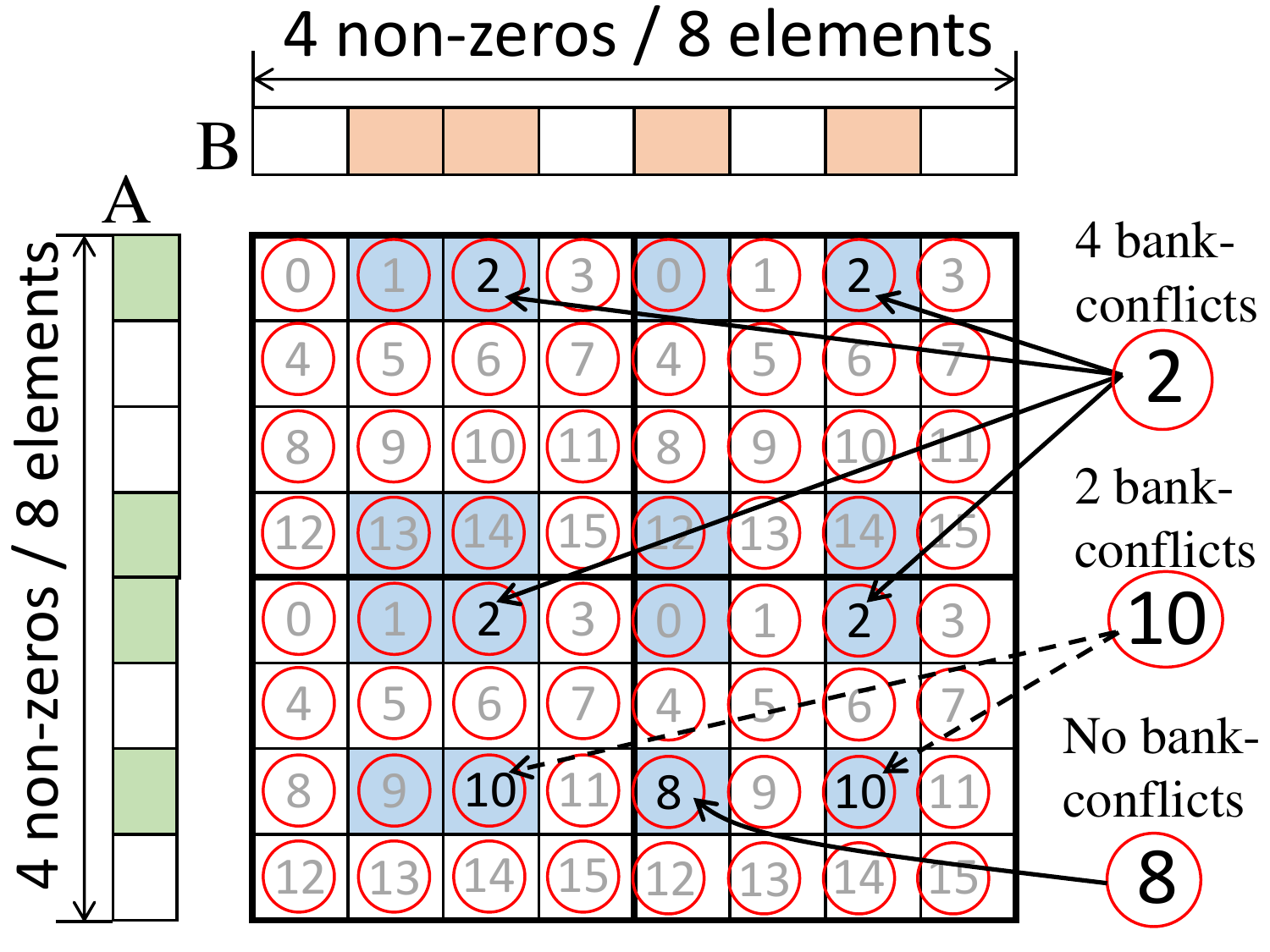}\label{fig:accum_sparse}
}
\caption{\small Memory access pattern in the accumulation buffer.}
\vspace*{-0.05in}
\label{fig:accum_port}
\end{figure}

The accumulation buffer has two modes, a \textit{dense} mode, and a \textit{sparse} mode. In dense mode, the accumulation buffer configures each read/write port directly connected to each output from FEOP units. In sparse mode, a large amount of partial matrix is generated (e.g., $32\times32$ FP32 for the warp-tile in SpWMMA). We extend the accumulation buffer to a multi-bank memory of 4 KByte ($32\times32\times4$ Bytes). Furthermore, the gather-accumulate-scatter method, discussed in Section~\ref{sec:algo:accumulate}, requires random access to multiple banks. We design an operand collector to schedule bank reads and writes to optimize the effective bandwidth.

\paragraph{Dense mode} Figure~\ref{fig:accum_dense} shows an example of the FEOPs' outputs memory access to accumulation buffer port. For simplicity, we use a $4\times4$ example. Since one OHMMA instruction is issued per cycle, the accumulation buffer uses 16 ports (e.g., the numbers in circles) for each FEOP output (e.g., elements in the blue matrix). 

\begin{figure}[t]
\centering
\subfloat[][W/o operand collector.]{\includegraphics[width=0.35\linewidth]{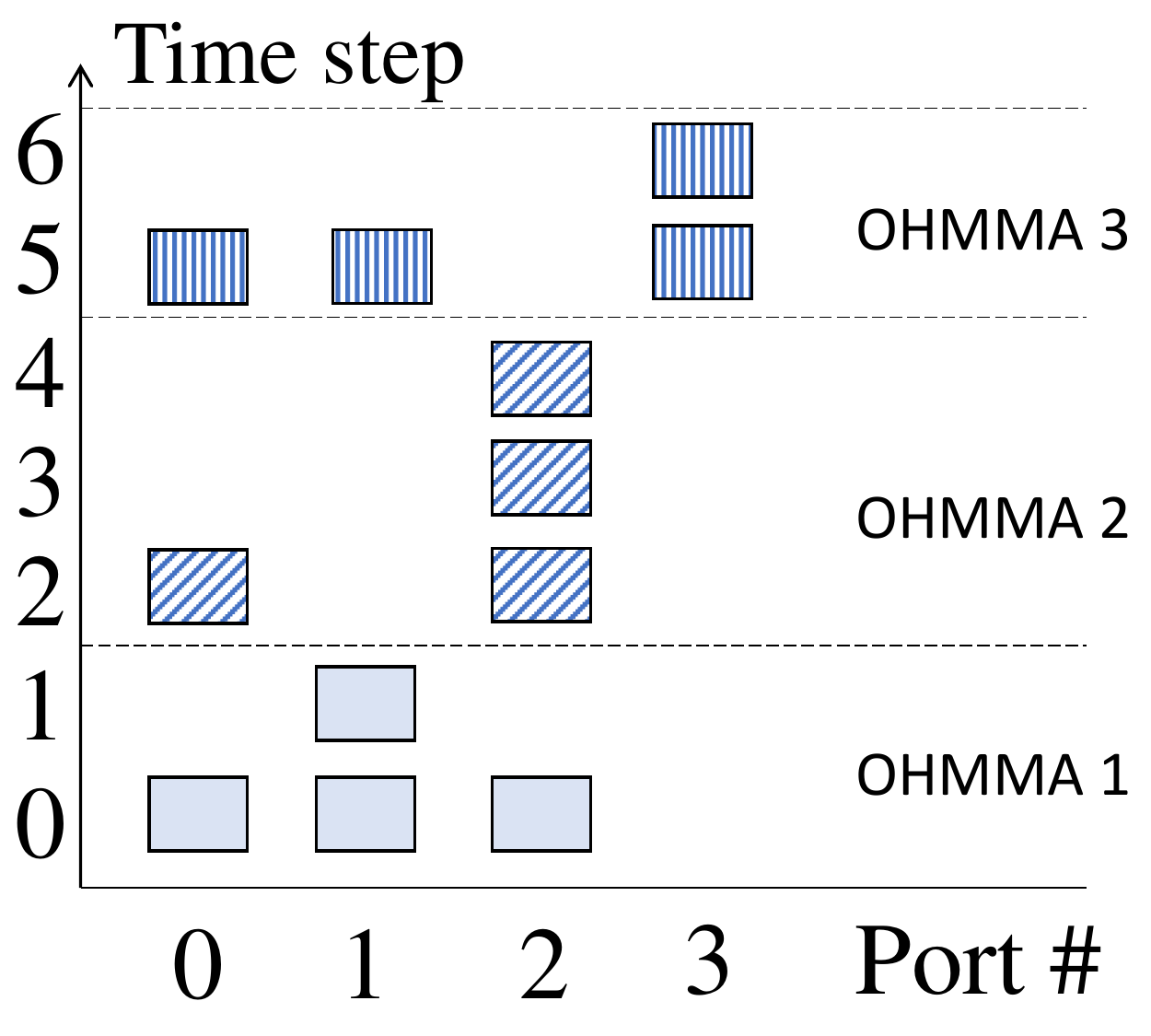}\label{fig:oper_collect_1}}
\subfloat[][With operand collector]{
\includegraphics[width=0.4\linewidth]{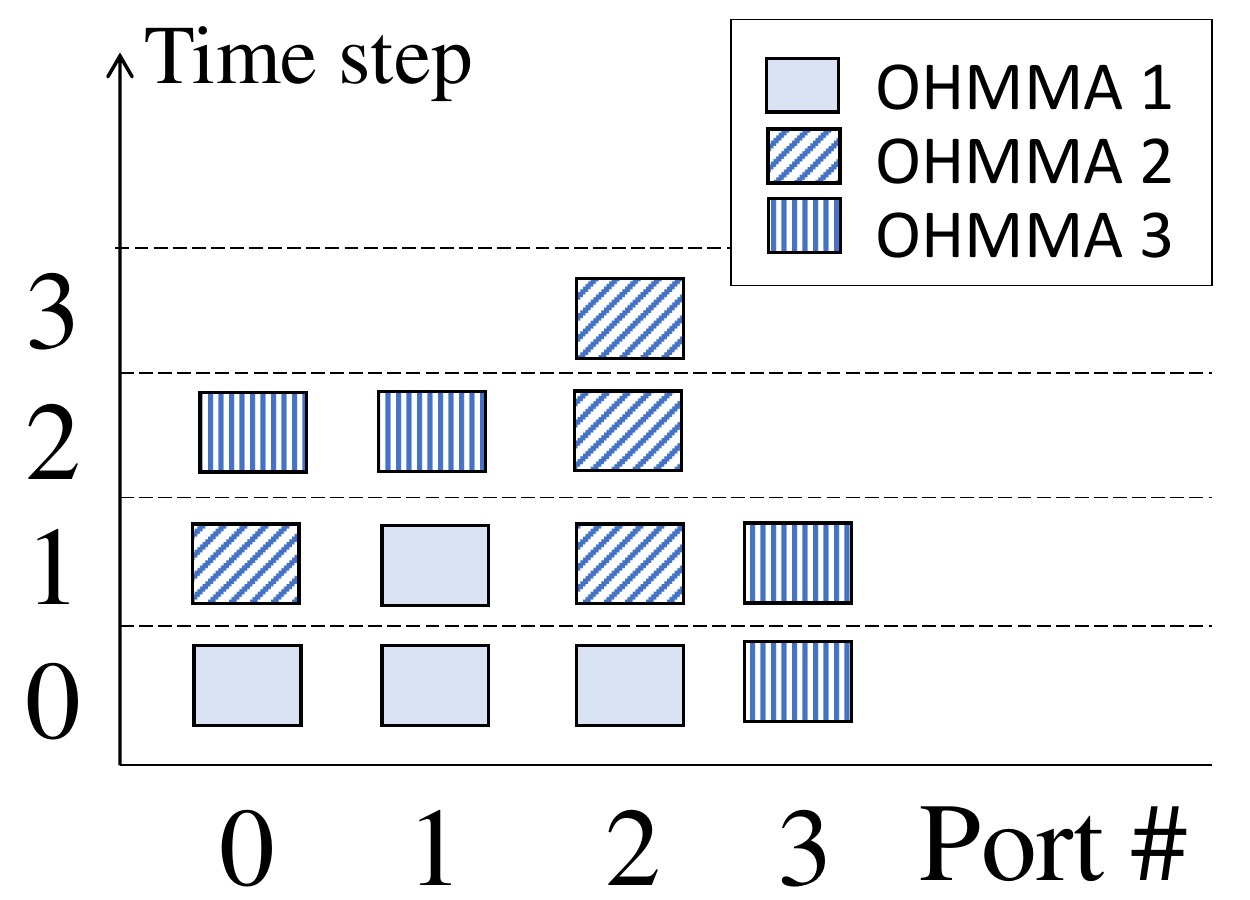}\label{fig:oper_collect_2}
}
\caption{\small Memory access schedule optimization via operand collector.} 
\label{fig:oper_collect}
\end{figure}
\begin{figure}[t]
\centering
\includegraphics[width=0.8\linewidth]{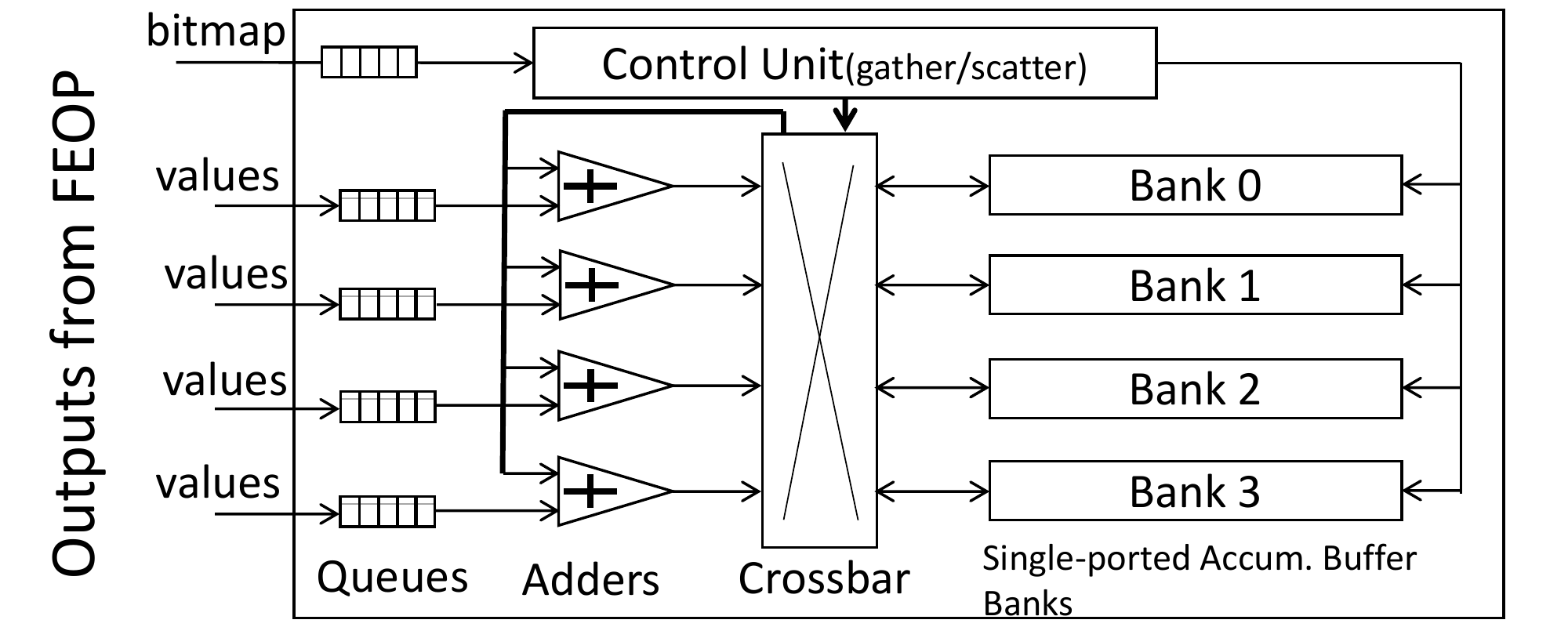}
\caption{\small Accumulation buffer design.}
\vspace*{-0.2in}
\label{fig:oper_col_diagram}
\end{figure}

\paragraph{Sparse mode} Figure~\ref{fig:accum_sparse} shows an example of accumulation buffer port memory access pattern when FEOP is running the \textit{sparse} matrix multiplication (e.g., SpWMMA in Figure~\ref{fig:spwmma_exe}). We assume an $8\times8\times1$ warp tile with both input vectors being 50\% sparse. Under this setting, OHMMA still generates 16 outputs per cycle. But the accumulation of these outputs may cause many bank conflicts as they are randomly distributed across the partial matrix, as shown in Figure~\ref{fig:accum_sparse}.

To solve this issue, we propose to integrate a small operand collector in our accumulation buffer for better memory bandwidth utilization. Operand collector \cite{operandcollector} is a technique used in NVIDIA's GPU micro-architecture to overlap the reading on the source operands from register file banks among multiple instructions. Figure~\ref{fig:oper_collect} shows an example of our accumulation buffer's memory access schedules with four operands per cycle on four ports with(out) operand collectors. The operand collector can significantly increase memory throughput by combining non-conflict memory accesses from different instructions.

Figure~\ref{fig:oper_col_diagram} shows an overall design of our accumulation buffer integrated with the aforementioned operand collector. It can support both dense and sparse outer product. The sparse mode is automatically turned on with the SpWMMA API.

\section{Evaluation}
\label{sec:exp}

We conduct comprehensive experiments with various micro benchmarks and DNN models to evaluate our software and hardware design, focusing on 1) how effective the bitmap-based im2col can reduce the decoding overhead; 2) how much performance our SpGEMM can improve upon a variety of sparsity ratios; 3) how much we can speed up the inference of diverse neural-network layers; and 4) how much hardware overhead is introduced in terms of hardware area. Evaluation results show that our design achieves significant performance improvements of up to one order of magnitude compared to the baselines and imposes small hardware overhead.
\subsection{Experimental Setup and Methodology}
\textbf{Simulation Platform} We use Accel-Sim~\cite{khairy2020accel}, a cycle-accurate simulator based on GPGPU-Sim~\cite{DBLP:conf/ispass/LewSPCZSNGSRA19}, to evaluate our design. The simulator provides flexible front-end architecture, optimizes cache and shared memory models, and improves simulation accuracy significantly compared with previous generations. We model a Tesla V100 GPU\cite{v100} on the simulator. To support SpWMMA instructions, we implement a cycle-accurate tensor core model based on our hardware design in Section~\ref{sec:arch}. In addition, we extend the simulator front-end to support our instruction extensions, as shown in Figure~\ref{fig:spwmma_exe}.

\textbf{Baselines} We choose CUTLASS~\cite{cutlass} and cuDNN~\cite{chetlur2014cudnn} as dense GEMM and convolution baselines, respectively. CUTLASS is an open-sourced GEMM library that achieves high performance comparable with cuBLAS~\cite{cublas}. cuDNN~\cite{chetlur2014cudnn} is a vendor-optimized, widely used library for DNN  acceleration. For our SpGEMM and SpCONV algorithms, we compare with two baselines: the vendor-optimized sparse matrix library cuSparse~\cite{naumov2010cusparse}, and the state-of-the-art research work of Sparse Tensor Core~\cite{sparsetensor}. For fair comparisons, our SpGEMM and SpCONV implementations build on the same loop tiling and software computation pipeline as CUTLASS~\cite{cutlass}.



\textbf{DNN Models and Pruning} We evaluate our algorithms using various types of DNN models, including 
1) three widely-used CNN models: VGG-16~\cite{DBLP:journals/corr/SimonyanZ14a}, ResNet-18~\cite{resnet}, and \hl{Mask R-CNN~\cite{MASKRCNN}}; 
2) one RNN model for word-level language modeling with a 2-layer LSTM encoder and a 4-layer LSTM decoder that was also used in Sparse Tensor Core\cite{sparsetensor} and we use the same configuration; 
and 3) BERT-base\cite{DBLP:journals/corr/abs-1810-04805} encoder, a representative and well-known attention-based model.

\begin{figure*}[!ht]
\centering
\begin{minipage}[ht]{\linewidth}
  \centering
  \includegraphics[width=1\linewidth]{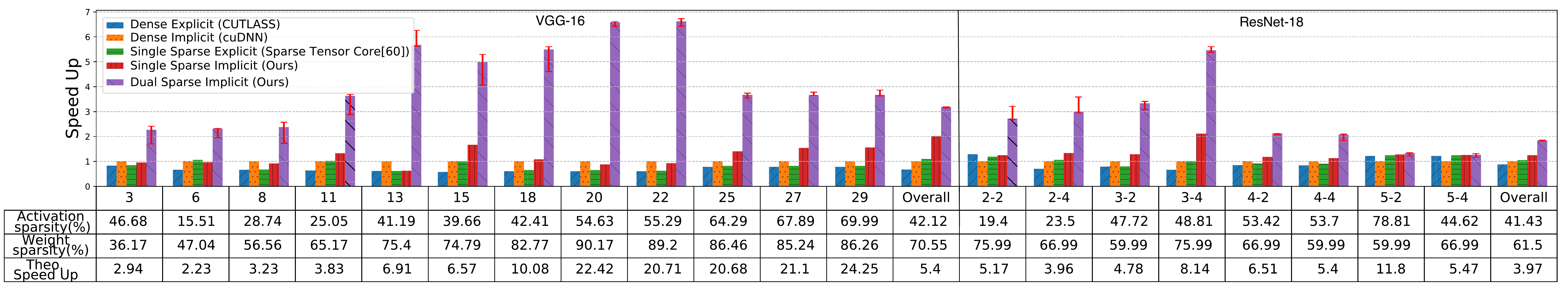}
\end{minipage}
\begin{minipage}[ht]{\linewidth}
 \centering
  \includegraphics[width=1\linewidth]{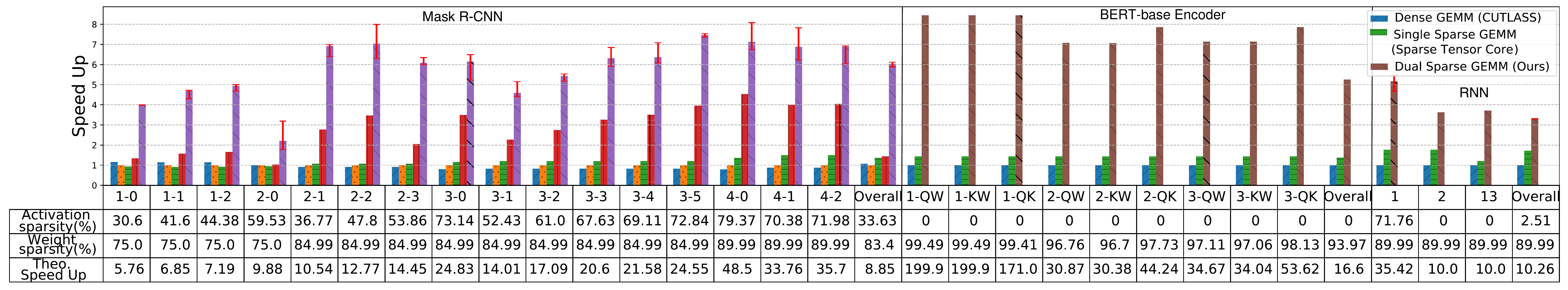}
  \caption{\small Model-inference performance comparison for different layers in the five DNN models. \hl{Note that the theoretical speedup is a loose upper bound, e.g., $100\times$ speedup on $99\%$ sparsity.}}
  \label{fig:speedup}
\end{minipage}
\vspace{-0.10in}
\end{figure*}

We fine-tune and prune the CNN models with Automated Gradual Pruner (AGP) \cite{DBLP:conf/iclr/ZhuG18} on \hl{Distiller\cite{DBLP:journals/corr/abs-1910-12232}}. \hl{We use the fine-pruned BERT-base encoder model \cite{BERT_PRUNE}\cite{huggingface} on the SQuAD task. We also fine-tune and prune the RNN model with AGP on Wikitext-2\cite{Wikitext} dataset. Unlike CNN models, BERT encoder and RNN models usually have high sparsity on only weights but not feature maps. Note that our work does not affect the model accuracy because we do not propose any new pruning algorithm. Table \ref{tab:setting} summarizes the sparse model accuracy, which is consistent with previous pruning works. The detailed layer-wise activation and weight sparsity ratios are listed in Figure~\ref{fig:speedup}.
}

\begin{table}[t]
\centering
\caption{\small \hl{Details of our evaluated sparse DNN model.}}
\label{tab:setting}
\begin{tabular}{|c||c|c|c|c|}
\hline
Models & Pruning Scheme & Dataset & Accuracy \\ \hline\hline
VGG-16 & \multirow{3}{*}{AGP\cite{DBLP:conf/iclr/ZhuG18}} & ImageNet & 88.86\% (top 5) \\ 
ResNet-18 &  & ImageNet & 86.46\% (top 5) \\ 
Mask R-CNN &  & COCO & 35.2 (AP) \\ \hline
BERT-base encoder & MP\cite{huggingface}\cite{BERT_PRUNE} & SQuAD & 83.3 (F1 score) \\ \hline
RNN & AGP & WikiText-2 & 85.7 (ppl) \\ \hline
\end{tabular}
\vspace{-0.1in}
\end{table}



\subsection{Performance of Bitmap-based Im2col}
We first evaluate the performance of our bitmap-based im2col, compared with dense im2col and CSR-encoded im2col. 
We compare against the CSR~\cite{sato1963techniques} as it is one of the most widely-used sparse matrix encoding methods.
We implement these three im2col algorithms based on PyTorch ATEN library\cite{paszke2019pytorch} and use a typical convolution layer from ResNet-18 to make the comparison. We measure the execution time of these algorithms and normalize the results over the dense im2col case. We tune different feature-map sparsity of 0\% - 99.9\% and show the results in Table~\ref{tab:im2col}. 

The results reveal that our bitmap-based im2col significantly outperforms CSR-encoded im2col across different sparsity ratios and is one order of magnitude faster when the sparsity ratio is less than 50\%. Only when the sparsity ratio is extremely high, e.g., 99.9\%, CSR-encoded im2col achieves a comparable (but still lower) performance with our bitmap-based im2col. This big discrepancy is attributed to the fact that CSR encoding introduces two additional data-dependent memory reads for each non-zero data access, while bitmap encoding compresses non-zero data offsets into bits that significantly reduce the operational intensity in im2col.



\begin{table}[t]
\centering
\caption{\small Normalized im2col time comparison using a typical convolution layer from ResNet-18 (feature map H/W=56, filter H/W=3, in/out channel=128) under different sparsity ratios.}
\label{tab:im2col}
\begin{tabular}{|c||c|c|c|c|c|c|}
\hline
Sparsity (\%)  & 0     & 25   & 50   & 75   & 99  & 99.9 \\ \hline
Dense Im2col  & 1     & 1    & 1    & 1    & 1   & 1    \\ \hline
CSR Im2col    & 101.3 & 67.1 & 45.2 & 14.5 & 4.7 & 1.2  \\ \hline
Bitmap Im2col & 8.31  & 6.87 & 4.73 & 2.5  & 1.5 & 1.1  \\ \hline
\end{tabular}
\end{table}

\begin{figure}[t]
 \centering
  \includegraphics[width=1\linewidth]{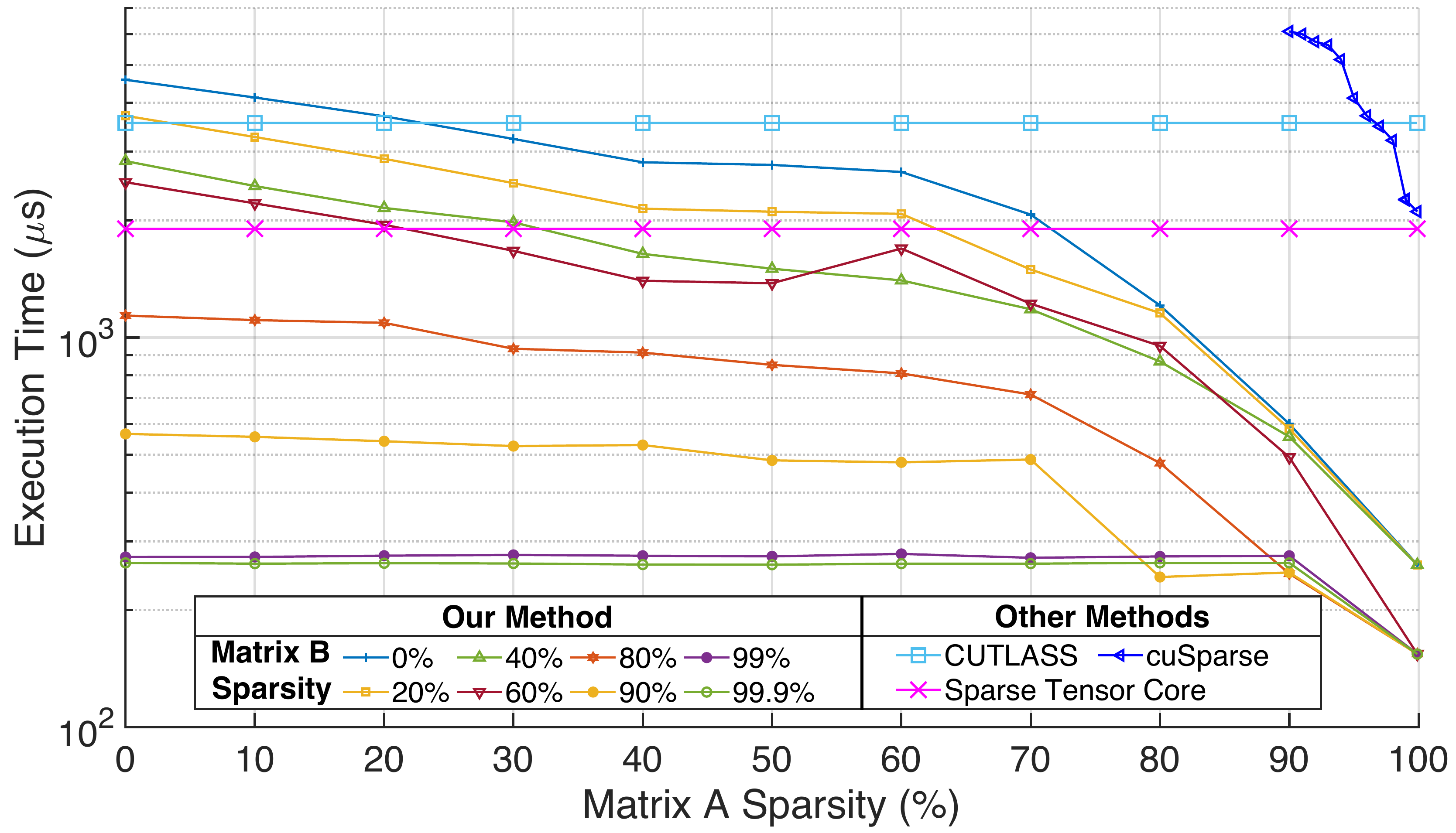}
  \caption{\small Performance comparison of SpGEMM on the CUTLASS baseline. cuSparse only outperforms the baseline when the sparsity is large($>$95\%). CSR-based Sparse Tensor Core cannot fully exploit dual-side sparsity. Our SpGEMM achieves a much higher speedup and also supports a very wide range of sparsity of matrices A and B.}
  \label{fig:exp1}
\end{figure}

\subsection{Performance of SpGEMM}

We then evaluate the performance of our SpGEMM, compared with CUTLASS, cuSparse, and Sparse Tensor Core~\cite{sparsetensor}. Among them, CUTLASS is the baseline of dense matrix multiplication. We measure the execution time of the multiplication of matrix A and matrix B (both are 4096$\times$4096) with various sparsity ratios. For cuSparse, we fix the sparsity of matrix B to 99\% and vary the sparsity of matrix A from only 90\% to 99.9\% because it is too slow when the sparsity of matrix A is less than 90\%. Figure~\ref{fig:exp1} shows the results, and we make the following observations. 

First, cuSparse is not applicable for accelerating sparse neural networks. Although matrix B already has a high sparsity of 99\%, cuSparse becomes faster than CUTLASS (i.e., the dense case) only when the sparsity of matrix A is higher than 95\%. Even when the sparsity of matrix A is as high as 99.9\%, it achieves a speedup of only 1.67$\times$. When the sparsity of matrix A is 90\%, it is 1.75 times slower than CUTLASS. Thus, cuSparse is only useful for matrix multiplication with extremely high sparsity, which rarely happens in practice.  

Second, Sparse Tensor Core~\cite{sparsetensor} has a fixed speedup of 1.86$\times$ over CUTLASS because it is designed  
to apply a fixed pruning ratio of 75\% (and thus the same sparsity ratio of 75\%) on only one sparse input matrix (matrix B in this experiment). 
As a result, it cannot take advantage of the other input matrix's sparsity, which significantly limits the acceleration ratio.

In contrast, our bitmap-based dual-side SpGEMM exploits the sparsity of both input and weight matrices and thus performs much better than cuSparse and Sparse Tensor Core. E.g., when the sparsity of matrix B is 99\%, we achieve a speedup of 13.4$\times$ over CUTLASS even when the sparsity of matrix A is 0; and if the sparsity of matrix A increases to 99.9\%, the speedup is as high as 23$\times$, which is 13.7$\times$ better than cuSparse under the same sparsity of matrices A and B (i.e., 99.9\% and 99\%, respectively). Furthermore, even when the sparsity of matrix B is 0, our SpGEMM becomes faster than CUTLASS when the sparsity of matrix A is higher than $\sim$25\%. 
Consequently, our SpGEMM achieves a significant speedup over the dense case for a wide range of sparsity (i.e., unless the sparsity ratios of \emph{both} matrix A and matrix B are lower than $\sim$25\%), making it highly useful for various SpMM.




\subsection{Performance of Real Neural-Network Inference}

Putting it together, we next evaluate the performance of real neural-network inference using the aforementioned five DNN models. For CNN models, we compare the performance in five cases: 1) \textit{Dense Explicit} is dense GEMM based on CUTLASS with explicit im2col; 
2) \textit{Dense Implicit} is dense GEMM with implicit im2col 
provided by cuDNN; 3) \textit{Single Sparse Explicit} is Sparse Tensor Core~\cite{sparsetensor} 
with explicit im2col; 
4) \textit{Single Sparse Implicit} only exploits weight matrix sparsity with our SpCONV; and 5) \textit{Dual Sparse Implicit} is our dual-side sparsity method with both feature map and weight sparsity. For BERT-base encoder and RNN models without im2col, we compare the performance in three cases: 1) \textit{Dense GEMM} based on CUTLASS; 2) \textit{Single Sparse GEMM} based on Sparse Tensor Core~\cite{sparsetensor}; and 3) \textit{Dual Sparse GEMM} which is our method.

\hl{Figure~\ref{fig:speedup} shows the layer-wise and full-model speedup of the five DNNs. We select a set of representative layers for brevity because the rest layers have the same shape.}
For CNN models, the speedup is normalized to \textit{Dense Implicit} which outperforms \textit{Dense Explicit} due to optimized im2col operation in convolution. \textit{Single Sparse Explicit}~\cite{sparsetensor} is also faster than \textit{Dense Explicit} by leveraging the sparsity of weight matrix, but is not always faster than \textit{Dense Implicit} with a speedup ranging from 0.78$\times$ to 1.74$\times$ (1.36$\times$ on average). Benefiting from our bitmap-based implicit im2col and sparse weight matrix, \textit{Single Sparse Implicit} is faster than \textit{Dense Implicit} in most cases, even it only takes advantage of weight matrix sparsity. It achieves an average speedup of \hl{1.92$\times$} ranging from \hl{0.63$\times$} to 4.5$\times$. By exploiting dual side sparsity and bitmap-based implicit im2col, our \textit{Dual Sparse Implicit} method significantly outperforms all the other methods and achieves a speedup of \hl{1.25$\times$-7.49$\times$ over \textit{Dense Implicit}}. The average speedup is 4.38$\times$ which is 2.22$\times$ higher than \textit{Single Sparse Explicit}~\cite{sparsetensor}. 

\hl{
Figure~\ref{fig:speedup} also shows that our method achieves a speedup close to the theoretical upper bound in some CONV layers.
A tight estimation of the upper bound is difficult because it depends on the non-zeros' distributions.
The small speedups for some layers (e.g., ResNet-18 layer 5-4) are due to their small sizes, where the performance is bound by data movement.
}

For BERT-base encoder and RNN models, the speedup is normalized to \textit{Dense GEMM}. \textit{Single Sparse GEMM}~\cite{sparsetensor} is always faster than \textit{Dense GEMM} but has a small speedup of only 1.20$\times$-1.77$\times$ (1.51$\times$ on average). Our method significantly outperforms \textit{Single Sparse GEMM} with a speed of 3.62$\times$-8.45$\times$. Our average speedup is 6.74$\times$, which is 3.46$\times$ higher than \textit{Single Sparse GEMM} because Sparse Tensor Core \cite{sparsetensor} only accelerates SpMM with a hard limit of 75\%, while the pruned BERT-base encoder model \cite{BERT_PRUNE} and RNN \cite{DBLP:conf/iclr/ZhuG18} has more than 90\% weight sparsity. 
\hl{Recall the example in Figure~\ref{fig:warp_speedup}, our work can go beyond the fixed-ratio limit due to our sparse tiling approach. For very sparse matrices, the proposed two-level bitmap encoding also helps because some empty warps are skipped as a whole, as shown in Figure~\ref{fig:two_level_con_3}.}




\subsection{Hardware Overhead}
Finally, we evaluate the hardware overhead \hl{and power consumption} of shared buffers and queues using CACTI 7~\cite{balasubramonian2017cacti} with 22~nm process technology and scale them to 12~nm\cite{stillmaker2017scaling}. We estimate Accumulation Operand Collector and Float Point Adders overheads {and energy consumption} in RTL implementation. As shown in Table~\ref{tab:my-table}, our design introduces a total hardware overhead of 12.846 $mm^2$, which is 1.5\% of the whole V100 die area of 815 $mm^2$, \hl{and it consumes an additional 3.89~W that is 1.6\% of V100's 250~W TDP.}


\begin{table}[ht]
\centering
\caption{\hl{Area and power overhead estimation.}}
\label{tab:my-table}
\begin{tabular}{|c|c|c|}
\hline
Module Name                    & \begin{tabular}[c]{@{}c@{}}Area Overhead\\ ($mm^2$, 12 nm)\end{tabular} & \begin{tabular}[c]{@{}c@{}}\hl{Power Consumption}\\ \hl{(W, 12 nm)}\end{tabular} \\ \hline \hline
Float Point Adders             & 0.121                                                                   & \hl{2.35}                                                                   \\ \hline
Accumulation Operand Collector & 1.51                                                                    & \hl{0.46}                                                                   \\ \hline
Shared Accumulation Buffer     & 11.215                                                                  & \hl{1.08}                                                                   \\ \hline \hline
Total overhead on V100         & 12.846 (1.5\%)                                                          & \hl{3.89 (1.60\%)}                                                                   \\ \hline






\end{tabular}
\vspace{-0.20in}
\end{table}
\section{Conclusion}
In this paper, for the first time, we demonstrate the feasibility of achieving a meaningful speedup for both SpGEMM and SpCONV on GPU Tensor Core with minimal hardware extension. The key insight is combing outer product of matrix multiplication and bitmap-base sparse encoding to fully leverage dual-side sparsity for highly-efficient GEMM and implicit im2col. Our design supports a wide range of sparsity ratios and outperforms state-of-the-art baselines by up to one order of magnitude with negligible hardware overhead, shedding light for the next performance breakthrough of future GPUs.

\vspace*{0.1cm}\noindent\textbf{Acknowledgements}\hspace*{0.2cm}
We thank the anonymous reviews for their thoughtful comments and suggestions.
The contribution of Jingwen Leng to this work was supported by the National Natural Science Foundation of China (NSFC) grant 62072297.

\bibliographystyle{IEEEtranS}
\bibliography{references}

\begin{thebibliography}{10}
\providecommand{\url}[1]{#1}
\csname url@samestyle\endcsname
\providecommand{\newblock}{\relax}
\providecommand{\bibinfo}[2]{#2}
\providecommand{\BIBentrySTDinterwordspacing}{\spaceskip=0pt\relax}
\providecommand{\BIBentryALTinterwordstretchfactor}{4}
\providecommand{\BIBentryALTinterwordspacing}{\spaceskip=\fontdimen2\font plus
\BIBentryALTinterwordstretchfactor\fontdimen3\font minus
  \fontdimen4\font\relax}
\providecommand{\BIBforeignlanguage}[2]{{%
\expandafter\ifx\csname l@#1\endcsname\relax
\typeout{** WARNING: IEEEtranS.bst: No hyphenation pattern has been}%
\typeout{** loaded for the language `#1'. Using the pattern for}%
\typeout{** the default language instead.}%
\else
\language=\csname l@#1\endcsname
\fi
#2}}
\providecommand{\BIBdecl}{\relax}
\BIBdecl

\bibitem{agarap2018deep}
A.~F. Agarap, ``Deep learning using rectified linear units (relu),''
  \emph{CoRR}, vol. abs/1803.08375, 2018.

\bibitem{aimar2018nullhop}
A.~Aimar, H.~Mostafa, E.~Calabrese, A.~Rios-Navarro, R.~Tapiador-Morales, I.-A.
  Lungu, M.~B. Milde, F.~Corradi, A.~Linares-Barranco, S.-C. Liu \emph{et~al.},
  ``Nullhop: A flexible convolutional neural network accelerator based on
  sparse representations of feature maps,'' \emph{IEEE transactions on neural
  networks and learning systems}, 2018.

\bibitem{cnvlutin}
J.~Albericio, P.~Judd, T.~Hetherington, T.~Aamodt, N.~E. Jerger, and
  A.~Moshovos, ``Cnvlutin: Ineffectual-neuron-free deep neural network
  computing,'' \emph{ACM SIGARCH Computer Architecture News}, 2016.

\bibitem{balasubramonian2017cacti}
R.~Balasubramonian, A.~B. Kahng, N.~Muralimanohar, A.~Shafiee, and V.~Srinivas,
  ``Cacti 7: New tools for interconnect exploration in innovative off-chip
  memories,'' \emph{ACM Transactions on Architecture and Code Optimization
  (TACO)}, vol.~14, no.~2, pp. 1--25, 2017.

\bibitem{Wikitext}
J.~Bradbury, S.~Merity, C.~Xiong, and R.~Socher, ``Quasi-recurrent neural
  networks,'' in \emph{International Conference on Learning Representations
  (ICLR)}, 2017.

\bibitem{seernet}
S.~Cao, L.~Ma, W.~Xiao, C.~Zhang, Y.~Liu, L.~Zhang, L.~Nie, and Z.~Yang,
  ``Seernet: Predicting convolutional neural network feature-map sparsity
  through low-bit quantization,'' in \emph{Proceedings of the IEEE Conference
  on Computer Vision and Pattern Recognition}, 2019, pp. 11\,216--11\,225.

\bibitem{fpga19}
S.~Cao, C.~Zhang, Z.~Yao, W.~Xiao, L.~Nie, D.~Zhan, Y.~Liu, M.~Wu, and
  L.~Zhang, ``Efficient and effective sparse lstm on fpga with bank-balanced
  sparsity,'' in \emph{Proceedings of the ACM/SIGDA International Symposium on
  Field-Programmable Gate Arrays}, 2019, pp. 63--72.

\bibitem{eyeriss}
Y.-H. Chen, T.~Krishna, J.~S. Emer, and V.~Sze, ``Eyeriss: An energy-efficient
  reconfigurable accelerator for deep convolutional neural networks,''
  \emph{IEEE journal of solid-state circuits}, 2016.

\bibitem{chetlur2014cudnn}
S.~Chetlur, C.~Woolley, P.~Vandermersch, J.~Cohen, J.~Tran, B.~Catanzaro, and
  E.~Shelhamer, ``cudnn: Efficient primitives for deep learning,'' \emph{arXiv
  preprint arXiv:1410.0759}, 2014.

\bibitem{choquette2020nvidia}
J.~Choquette and W.~Gandhi, ``Nvidia a100 gpu: Performance \& innovation for
  gpu computing,'' in \emph{2020 IEEE Hot Chips 32 Symposium (HCS)}.\hskip 1em
  plus 0.5em minus 0.4em\relax IEEE Computer Society, 2020, pp. 1--43.

\bibitem{volta}
J.~Choquette, O.~Giroux, and D.~Foley, ``Volta: Performance and
  programmability,'' \emph{Ieee Micro}, vol.~38, no.~2, pp. 42--52, 2018.

\bibitem{cong2014minimizing}
J.~Cong and B.~Xiao, ``Minimizing computation in convolutional neural
  networks,'' in \emph{International conference on artificial neural
  networks}.\hskip 1em plus 0.5em minus 0.4em\relax Springer, 2014, pp.
  281--290.

\bibitem{DBLP:journals/corr/abs-1810-04805}
J.~Devlin, M.~Chang, K.~Lee, and K.~Toutanova, ``{BERT:} pre-training of deep
  bidirectional transformers for language understanding,'' \emph{CoRR}, vol.
  abs/1810.04805, 2018.

\bibitem{dong2017more}
X.~Dong, J.~Huang, Y.~Yang, and S.~Yan, ``More is less: A more complicated
  network with less inference complexity,'' in \emph{IEEE Conference on
  Computer Vision and Pattern Recognition}, 2017.

\bibitem{figurnov2017spatially}
M.~Figurnov, M.~D. Collins, Y.~Zhu, L.~Zhang, J.~Huang, D.~Vetrov, and
  R.~Salakhutdinov, ``Spatially adaptive computation time for residual
  networks,'' in \emph{Proceedings of the IEEE Conference on Computer Vision
  and Pattern Recognition}, 2017, pp. 1039--1048.

\bibitem{ptolemy_micro20}
Y.~{Gan}, Y.~{Qiu}, J.~{Leng}, M.~{Guo}, and Y.~{Zhu}, ``{Ptolemy: Architecture
  Support for Robust Deep Learning},'' in \emph{2020 53rd Annual IEEE/ACM
  International Symposium on Microarchitecture (MICRO)}, 2020.

\bibitem{sparten}
A.~Gondimalla, N.~Chesnut, M.~Thottethodi, and T.~Vijaykumar, ``Sparten: A
  sparse tensor accelerator for convolutional neural networks,'' in
  \emph{Proceedings of the 52nd Annual IEEE/ACM International Symposium on
  Microarchitecture}, 2019, pp. 151--165.

\bibitem{lstm}
K.~Greff, R.~K. Srivastava, J.~Koutn{\'\i}k, B.~R. Steunebrink, and
  J.~Schmidhuber, ``Lstm: A search space odyssey,'' \emph{IEEE transactions on
  neural networks and learning systems}, vol.~28, no.~10, pp. 2222--2232, 2016.

\bibitem{guan2020far}
Y.~Guan, J.~Leng, C.~Li, Q.~Chen, and M.~Guo, ``{How Far Does BERT Look
  At:Distance-based Clustering and Analysis of BERT$'$s Attention},''
  \emph{arXiv preprint arXiv:2011.00943}, 2020.

\bibitem{tw_sc20}
C.~{Guo}, B.~Y. {Hsueh}, J.~{Leng}, Y.~{Qiu}, Y.~{Guan}, Z.~{Wang}, X.~{Jia},
  X.~{Li}, M.~{Guo}, and Y.~{Zhu}, ``Accelerating sparse dnn models without
  hardware-support via tile-wise sparsity,'' in \emph{International Conference
  for High Performance Computing, Networking, Storage and Analysis}, 2020.

\bibitem{sma_dac20}
C.~Guo, Y.~Zhou, J.~Leng, Y.~Zhu, Z.~Du, Q.~Chen, C.~Li, B.~Yao, and M.~Guo,
  ``{Balancing Efficiency and Flexibility for DNN Acceleration via Temporal
  GPU-Systolic Array Integration},'' in \emph{Proceedings of the Design
  Automation Conference}, 2020.

\bibitem{han2017ese}
S.~Han, J.~Kang, H.~Mao, Y.~Hu, X.~Li, Y.~Li, D.~Xie, H.~Luo, S.~Yao, Y.~Wang
  \emph{et~al.}, ``Ese: Efficient speech recognition engine with sparse lstm on
  fpga,'' in \emph{Proceedings of the 2017 ACM/SIGDA International Symposium on
  Field-Programmable Gate Arrays}, 2017, pp. 75--84.

\bibitem{eie}
S.~Han, X.~Liu, H.~Mao, J.~Pu, A.~Pedram, M.~A. Horowitz, and W.~J. Dally,
  ``Eie: efficient inference engine on compressed deep neural network,''
  \emph{ACM SIGARCH Computer Architecture News}, 2016.

\bibitem{han2015learning}
S.~Han, J.~Pool, J.~Tran, and W.~Dally, ``Learning both weights and connections
  for efficient neural network,'' in \emph{Advances in Neural Information
  Processing Systems}, 2015, pp. 1135--1143.

\bibitem{hazelwood2018applied}
K.~Hazelwood, S.~Bird, D.~Brooks, S.~Chintala, U.~Diril, D.~Dzhulgakov,
  M.~Fawzy, B.~Jia, Y.~Jia, A.~Kalro \emph{et~al.}, ``Applied machine learning
  at facebook: A datacenter infrastructure perspective,'' in \emph{2018 IEEE
  International Symposium on High Performance Computer Architecture
  (HPCA)}.\hskip 1em plus 0.5em minus 0.4em\relax IEEE, 2018, pp. 620--629.

\bibitem{MASKRCNN}
K.~He, G.~Gkioxari, P.~Doll{\'a}r, and R.~Girshick, ``Mask r-cnn,'' in
  \emph{IEEE international conference on computer vision}, 2017.

\bibitem{resnet}
K.~He, X.~Zhang, S.~Ren, and J.~Sun, ``Deep residual learning for image
  recognition,'' in \emph{Proceedings of the IEEE conference on computer vision
  and pattern recognition}, 2016, pp. 770--778.

\bibitem{hegde2019extensor}
K.~Hegde, H.~Asghari-Moghaddam, M.~Pellauer, N.~Crago, A.~Jaleel, E.~Solomonik,
  J.~Emer, and C.~W. Fletcher, ``Extensor: An accelerator for sparse tensor
  algebra,'' in \emph{Proceedings of the 52nd Annual IEEE/ACM International
  Symposium on Microarchitecture}, 2019, pp. 319--333.

\bibitem{extensor}
------, ``Extensor: An accelerator for sparse tensor algebra,'' in
  \emph{Proceedings of the 52nd Annual IEEE/ACM International Symposium on
  Microarchitecture}, 2019, pp. 319--333.

\bibitem{huggingface}
\BIBentryALTinterwordspacing
Huggingface. [Online]. Available:
  \url{https://github.com/huggingface/block_movement_pruning#fine-pruned-models}
\BIBentrySTDinterwordspacing

\bibitem{ioannou2017deep}
Y.~Ioannou, D.~Robertson, R.~Cipolla, and A.~Criminisi, ``Deep roots: Improving
  cnn efficiency with hierarchical filter groups,'' in \emph{IEEE Conference on
  Computer Vision and Pattern Recognition}, 2017.

\bibitem{wavernn}
N.~Kalchbrenner, E.~Elsen, K.~Simonyan, S.~Noury, N.~Casagrande, E.~Lockhart,
  F.~Stimberg, A.~Oord, S.~Dieleman, and K.~Kavukcuoglu, ``Efficient neural
  audio synthesis,'' in \emph{International Conference on Machine
  Learning}.\hskip 1em plus 0.5em minus 0.4em\relax PMLR, 2018, pp. 2410--2419.

\bibitem{khairy2020accel}
M.~Khairy, Z.~Shen, T.~M. Aamodt, and T.~G. Rogers, ``Accel-sim: an extensible
  simulation framework for validated gpu modeling,'' in \emph{2020 ACM/IEEE
  47th Annual International Symposium on Computer Architecture (ISCA)}.\hskip
  1em plus 0.5em minus 0.4em\relax IEEE, 2020, pp. 473--486.

\bibitem{kong2017ron}
T.~Kong, F.~Sun, A.~Yao, H.~Liu, M.~Lu, and Y.~Chen, ``Ron: Reverse connection
  with objectness prior networks for object detection,'' in \emph{IEEE
  Conference on Computer Vision and Pattern Recognition}, 2017.

\bibitem{lavin2016fast}
A.~Lavin and S.~Gray, ``Fast algorithms for convolutional neural networks,'' in
  \emph{Proceedings of the IEEE Conference on Computer Vision and Pattern
  Recognition}, 2016, pp. 4013--4021.

\bibitem{DBLP:conf/ispass/LewSPCZSNGSRA19}
J.~Lew \emph{et~al.}, ``Analyzing machine learning workloads using a detailed
  {GPU} simulator,'' in \emph{{IEEE} International Symposium on Performance
  Analysis of Systems and Software, {ISPASS}}.\hskip 1em plus 0.5em minus
  0.4em\relax {IEEE}, 2019, pp. 151--152.

\bibitem{liu2015sparse}
B.~Liu, M.~Wang, H.~Foroosh, M.~Tappen, and M.~Pensky, ``Sparse convolutional
  neural networks,'' in \emph{Proceedings of the IEEE Conference on Computer
  Vision and Pattern Recognition}, 2015, pp. 806--814.

\bibitem{operandcollector}
S.~Liu, J.~E. Lindholm, M.~Y. Siu, B.~W. Coon, and S.~F. Oberman, ``Operand
  collector architecture,'' Nov.~16 2010, uS Patent 7,834,881.

\bibitem{liu2018efficient}
X.~Liu, J.~Pool, S.~Han, and W.~J. Dally, ``Efficient sparse-winograd
  convolutional neural networks,'' in \emph{International Conference on
  Learning Representations}, 2018.

\bibitem{lu2018spwa}
L.~Lu and Y.~Liang, ``Spwa: an efficient sparse winograd convolutional neural
  networks accelerator on fpgas,'' in \emph{2018 55th ACM/ESDA/IEEE Design
  Automation Conference (DAC)}.\hskip 1em plus 0.5em minus 0.4em\relax IEEE,
  2018, pp. 1--6.

\bibitem{DBLP:conf/emnlp/LuongPM15}
T.~Luong, H.~Pham, and C.~D. Manning, ``Effective approaches to attention-based
  neural machine translation,'' in \emph{The Conference on Empirical Methods in
  Natural Language Processing}, L.~M{\`{a}}rquez, C.~Callison{-}Burch, J.~Su,
  D.~Pighin, and Y.~Marton, Eds., 2015.

\bibitem{mishra2021accelerating}
A.~Mishra, J.~A. Latorre, J.~Pool, D.~Stosic, D.~Stosic, G.~Venkatesh, C.~Yu,
  and P.~Micikevicius, ``Accelerating sparse deep neural networks,'' 2021.

\bibitem{naumov2010cusparse}
M.~Naumov, L.~Chien, P.~Vandermersch, and U.~Kapasi, ``Cusparse library,'' in
  \emph{GPU Technology Conference}, 2010.

\bibitem{cublas}
Nvidia, ``Cublas library,'' \emph{NVIDIA Corporation, Santa Clara}, 2008.

\bibitem{a100}
------, ``Nvidia a100 tensor core architecture,'' in \emph{Technical
  report}.\hskip 1em plus 0.5em minus 0.4em\relax NVIDIA, 2020.

\bibitem{cutlass}
C.~Nvidia, ``Cutlass library,'' \emph{NVIDIA Corporation, Santa Clara,
  California}, vol.~15, no.~27, p.~31, 2008.

\bibitem{v100}
T.~NVIDIA, ``V100 gpu architecture. the world’s most advanced data center
  gpu. version wp-08608-001\_v1. 1,'' \emph{NVIDIA. Aug}, p. 108, 2017.

\bibitem{outerspace}
S.~Pal, J.~Beaumont, D.-H. Park, A.~Amarnath, S.~Feng, C.~Chakrabarti, H.-S.
  Kim, D.~Blaauw, T.~Mudge, and R.~Dreslinski, ``Outerspace: An outer product
  based sparse matrix multiplication accelerator,'' in \emph{2018 IEEE
  International Symposium on High Performance Computer Architecture
  (HPCA)}.\hskip 1em plus 0.5em minus 0.4em\relax IEEE, 2018, pp. 724--736.

\bibitem{scnn}
A.~Parashar, M.~Rhu, A.~Mukkara, A.~Puglielli, R.~Venkatesan, B.~Khailany,
  J.~Emer, S.~W. Keckler, and W.~J. Dally, ``Scnn: An accelerator for
  compressed-sparse convolutional neural networks,'' \emph{ACM SIGARCH Computer
  Architecture News}, vol.~45, no.~2, pp. 27--40, 2017.

\bibitem{paszke2019pytorch}
A.~Paszke, S.~Gross, F.~Massa, A.~Lerer, J.~Bradbury, G.~Chanan, T.~Killeen,
  Z.~Lin, N.~Gimelshein, L.~Antiga \emph{et~al.}, ``Pytorch: An imperative
  style, high-performance deep learning library,'' in \emph{Advances in neural
  information processing systems}, 2019, pp. 8026--8037.

\bibitem{path_cvpr19}
Y.~Qiu, J.~Leng, C.~Guo, Q.~Chen, C.~Li, M.~Guo, and Y.~Zhu, ``{Adversarial
  Defense Through Network Profiling Based Path Extraction},'' in
  \emph{Proceedings of the IEEE/CVF Conference on Computer Vision and Pattern
  Recognition (CVPR)}, June 2019.

\bibitem{raihan2019modeling}
M.~A. Raihan, N.~Goli, and T.~M. Aamodt, ``Modeling deep learning accelerator
  enabled gpus,'' in \emph{International Symposium on Performance Analysis of
  Systems and Software (ISPASS)}.\hskip 1em plus 0.5em minus 0.4em\relax IEEE,
  2019, pp. 79--92.

\bibitem{ren2018sbnet}
M.~Ren, A.~Pokrovsky, B.~Yang, and R.~Urtasun, ``Sbnet: Sparse blocks network
  for fast inference,'' in \emph{Proceedings of the IEEE Conference on Computer
  Vision and Pattern Recognition}, 2018, pp. 8711--8720.

\bibitem{BERT_PRUNE}
\BIBentryALTinterwordspacing
V.~Sanh, T.~Wolf, and A.~M. Rush, ``Movement pruning: Adaptive sparsity by
  fine-tuning,'' in \emph{Advances in Neural Information Processing Systems},
  2020. [Online]. Available:
  \url{https://papers.nips.cc/paper/2020/file/eae15aabaa768ae4a5993a8a4f4fa6e4-Paper.pdf}
\BIBentrySTDinterwordspacing

\bibitem{sato1963techniques}
N.~Sato and W.~Tinney, ``Techniques for exploiting the sparsity or the network
  admittance matrix,'' \emph{IEEE Transactions on Power Apparatus and Systems},
  vol.~82, no.~69, pp. 944--950, 1963.

\bibitem{shi2017speeding}
S.~Shi and X.~Chu, ``Speeding up convolutional neural networks by exploiting
  the sparsity of rectifier units,'' \emph{arXiv:1704.07724}, 2017.

\bibitem{DBLP:journals/corr/SimonyanZ14a}
K.~Simonyan and A.~Zisserman, ``Very deep convolutional networks for
  large-scale image recognition,'' in \emph{International Conference on
  Learning Representations (ICLR)}, Y.~Bengio and Y.~LeCun, Eds., 2015.

\bibitem{MatRaptor}
N.~Srivastava, H.~Jin, J.~Liu, D.~Albonesi, and Z.~Zhang, ``Matraptor: A
  sparse-sparse matrix multiplication accelerator based on row-wise product,''
  in \emph{2020 53rd Annual IEEE/ACM International Symposium on
  Microarchitecture (MICRO)}.\hskip 1em plus 0.5em minus 0.4em\relax IEEE,
  2020, pp. 766--780.

\bibitem{stillmaker2017scaling}
A.~Stillmaker and B.~Baas, ``Scaling equations for the accurate prediction of
  cmos device performance from 180 nm to 7 nm,'' \emph{Integration}, vol.~58,
  pp. 74--81, 2017.

\bibitem{varma2019dynamic}
G.~Varma, K.~Kothapalli \emph{et~al.}, ``Dynamic block sparse
  reparameterization of convolutional neural networks,'' in \emph{Proceedings
  of the IEEE International Conference on Computer Vision Workshops}, 2019, pp.
  0--0.

\bibitem{wen2016learning}
W.~Wen, C.~Wu, Y.~Wang, Y.~Chen, and H.~Li, ``Learning structured sparsity in
  deep neural networks,'' in \emph{Advances in Neural Information Processing
  Systems}, 2016, pp. 2074--2082.

\bibitem{9157343}
H.~{Yang}, S.~{Gui}, Y.~{Zhu}, and J.~{Liu}, ``Automatic neural network
  compression by sparsity-quantization joint learning: A constrained
  optimization-based approach,'' in \emph{IEEE Conference on Computer Vision
  and Pattern Recognition (CVPR)}, 2020.

\bibitem{yang2019energyconstrained}
H.~Yang, Y.~Zhu, and J.~Liu, ``Energy-constrained compression for deep neural
  networks via weighted sparse projection and layer input masking,'' in
  \emph{International Conference on Learning Representations}, 2018.

\bibitem{yang2019ecc}
------, ``Ecc: Platform-independent energy-constrained deep neural network
  compression via a bilinear regression model,'' in \emph{Proceedings of the
  IEEE/CVF Conference on Computer Vision and Pattern Recognition}, 2019, pp.
  11\,206--11\,215.

\bibitem{yao2014gpu}
Z.~Yao, V.~Gripon, and M.~Rabbat, ``A gpu-based associative memory using sparse
  neural networks,'' in \emph{International Conference on High Performance
  Computing \& Simulation (HPCS)}, 2014, pp. 688--692.

\bibitem{aaai19}
Z.~Yao, S.~Cao, W.~Xiao, C.~Zhang, and L.~Nie, ``Balanced sparsity for
  efficient dnn inference on gpu,'' in \emph{Proceedings of the AAAI Conference
  on Artificial Intelligence}, vol.~33, 2019, pp. 5676--5683.

\bibitem{yu2017scalpel}
J.~Yu, A.~Lukefahr, D.~Palframan, G.~Dasika, R.~Das, and S.~Mahlke, ``Scalpel:
  Customizing dnn pruning to the underlying hardware parallelism,'' \emph{ACM
  SIGARCH Computer Architecture News}, 2017.

\bibitem{caffeine}
C.~Zhang, G.~Sun, Z.~Fang, P.~Zhou, P.~Pan, and J.~Cong, ``Caffeine: Toward
  uniformed representation and acceleration for deep convolutional neural
  networks,'' \emph{IEEE Transactions on Computer-Aided Design of Integrated
  Circuits and Systems}, vol.~38, no.~11, pp. 2072--2085, 2018.

\bibitem{cambriconx}
S.~Zhang, Z.~Du, L.~Zhang, H.~Lan, S.~Liu, L.~Li, Q.~Guo, T.~Chen, and Y.~Chen,
  ``Cambricon-x: An accelerator for sparse neural networks,'' in
  \emph{International Symposium on Microarchitecture (MICRO)}, 2016.

\bibitem{sparch}
Z.~Zhang, H.~Wang, S.~Han, and W.~J. Dally, ``Sparch: Efficient architecture
  for sparse matrix multiplication,'' in \emph{International Symposium on High
  Performance Computer Architecture (HPCA)}, 2020.

\bibitem{cambricons}
X.~Zhou, Z.~Du, Q.~Guo, S.~Liu, C.~Liu, C.~Wang, X.~Zhou, L.~Li, T.~Chen, and
  Y.~Chen, ``Cambricon-s: Addressing irregularity in sparse neural networks
  through a cooperative software/hardware approach,'' in \emph{International
  Symposium on Microarchitecture (MICRO)}, 2018.

\bibitem{sparsetensor}
M.~Zhu, T.~Zhang, Z.~Gu, and Y.~Xie, ``Sparse tensor core: Algorithm and
  hardware co-design for vector-wise sparse neural networks on modern gpus,''
  in \emph{Proceedings of the 52nd Annual IEEE/ACM International Symposium on
  Microarchitecture}, 2019, pp. 359--371.

\bibitem{DBLP:conf/iclr/ZhuG18}
M.~Zhu and S.~Gupta, ``To prune, or not to prune: Exploring the efficacy of
  pruning for model compression,'' in \emph{6th International Conference on
  Learning Representations, {ICLR} 2018, Vancouver, BC, Canada, April 30 - May
  3, 2018, Workshop Track Proceedings}.

\bibitem{DBLP:journals/corr/abs-1910-12232}
N.~Zmora, G.~Jacob, L.~Zlotnik, B.~Elharar, and G.~Novik, ``Neural network
  distiller: {A} python package for {DNN} compression research,'' \emph{CoRR},
  vol. abs/1910.12232, 2019.

\end{thebibliography}

\end{document}